%% file: arxiv.tex
\newcommand{\accept}{\textit{ACCEPT}}
\newcommand{\hifl}{\textit{HIFLUGCS}}
\newcommand{\numobs}{317}
\newcommand{\numcluster}{239}
\newcommand{\expt}{9.86 Msec}
\newcommand{\dkna}{\ensuremath{(\kna^{\prime}-\kna)/\kna}}
\newcommand{\alphafs}{\ensuremath{\alpha = 1.21 \pm 0.39}}

\newcommand{\khunfs}{\ensuremath{\khun = 126 \pm 45 \ent}}
\newcommand{\alphaga}{\ensuremath{\alpha = 1.20 \pm 0.38}}

\newcommand{\khunga}{\ensuremath{\khun = 150 \pm 50 \ent}}
\newcommand{\alphagb}{\ensuremath{\alpha = 1.23 \pm 0.40}}

\newcommand{\khungb}{\ensuremath{\khun = 107 \pm 39 \ent}}
\newcommand{\centsrcnum}{\ensuremath{37}}

\newcommand{\kmma}{\ensuremath{K_1 = 17.8 \pm 6.6 \ent}}
\newcommand{\kmmb}{\ensuremath{K_2 = 154 \pm 52 \ent}}
\newcommand{\kmmc}{\ensuremath{124}}
\newcommand{\kmmd}{\ensuremath{109}} 
\newcommand{\kmme}{\ensuremath{p = 1.16\times10^{-7}}}
\newcommand{\kmmf}{\ensuremath{K_1 = 15.0\pm 5.0 \ent}}
\newcommand{\kmmg}{\ensuremath{K_2 = 129 \pm 45 \ent}}
\newcommand{\kmmh}{\ensuremath{89}}
\newcommand{\kmmi}{\ensuremath{136}}
\newcommand{\kmmj}{\ensuremath{p = 1.90\times10^{-13}}}

\newcommand{\hiflkmma}{\ensuremath{K_1 = 9.7 \pm 3.5 \ent}}
\newcommand{\hiflkmmb}{\ensuremath{K_2 = 131 \pm 46 \ent}}
\newcommand{\hiflkmmc}{\ensuremath{28}}
\newcommand{\hiflkmmd}{\ensuremath{31}}
\newcommand{\hiflkmme}{\ensuremath{p = 3.34\times10^{-3}}}
\newcommand{\hiflkmmf}{\ensuremath{K_1 = 10.5 \pm 3.4 \ent}}
\newcommand{\hiflkmmg}{\ensuremath{K_2 = 116 \pm 42 \ent}}
\newcommand{\hiflkmmh}{\ensuremath{21}}
\newcommand{\hiflkmmi}{\ensuremath{34}}
\newcommand{\hiflkmmj}{\ensuremath{p = 1.55\times10^{-5}}}
\newcommand{\tckmma}{\ensuremath{t_{c1} = 0.60 \pm 0.24 \Gyr}}
\newcommand{\tckmmb}{\ensuremath{t_{c2} = 6.23 \pm 2.19 \Gyr}}
\newcommand{\tckmmc}{\ensuremath{132}}
\newcommand{\tckmmd}{\ensuremath{101}}
\newcommand{\tckmme}{\ensuremath{p = 8.77\times10^{-7}}}

\documentclass{emulateapj}
\usepackage{apjfonts,graphicx,here,common,longtable,ifthen,amsmath,amssymb,natbib}
\usepackage[pagebackref,
  pdftitle={INTRACLUSTER MEDIUM ENTROPY PROFILES FOR A CHANDRA ARCHIVAL SAMPLE OF GALAXY CLUSTERS},
  pdfauthor={Kenneth W. Cavagnolo},
  pdfsubject={ApJ Supplement},
  pdfkeywords={clusters galaxies X-ray Chandra entropy},
  pdfproducer={LaTeX with hyperref},
  pdfcreator={LaTeX}
  pdfdisplaydoctitle=true,
  colorlinks=true,
  citecolor=blue,
  linkcolor=blue,
  urlcolor=blue]{hyperref}
\begin{document}
\title{INTRACLUSTER MEDIUM ENTROPY PROFILES\\FOR A CHANDRA ARCHIVAL SAMPLE OF GALAXY CLUSTERS}
\author{
  Kenneth. W. Cavagnolo\altaffilmark{1,2}, Megan
  Donahue\altaffilmark{1}, G. Mark Voit\altaffilmark{1}, and Ming
  Sun\altaffilmark{1,3}}
\altaffiltext{1}{Michigan State University, Department of Physics and
  Astronomy, East Lansing, MI, 48824-2320}
\altaffiltext{2}{University of Waterloo, Department of Physics and
  Astronomy, Waterloo, ON, N2L 2G1; kcavagno@scimail.uwaterloo.ca}
\altaffiltext{3}{University of Virginia, Department of Astronomy,
  Charlottesville, VA, 22904}
\shorttitle{ICM Entropy Profiles}
\shortauthors{K. W. Cavagnolo et al.}
\journalinfo{}
\slugcomment{Accepted for publication in The Astrophysical Journal Supplement Series}

\begin{abstract}
   We present radial entropy profiles of the intracluster medium (ICM)
   for a collection of \numcluster\ clusters taken from the
   \chandra\ X-ray Observatory's Data Archive. Entropy is of great
   interest because it controls ICM global properties and records the
   thermal history of a cluster. Entropy is therefore a useful
   quantity for studying the effects of feedback on the cluster
   environment and investigating any breakdown of cluster
   self-similarity. We find that most ICM entropy profiles are
   well-fit by a model which is a power-law at large radii and
   approaches a constant value at small radii: $K(r) = \kna + \khun
   (r/100 \kpc)^{\alpha}$, where \kna\ quantifies the typical excess
   of core entropy above the best fitting power-law found at larger
   radii. We also show that the \kna\ distributions of both the full
   archival sample and the primary \hifl\ sample of \citet{hiflugcs1}
   are bimodal with a distinct gap between $\kna \approx 30-50 \ent$
   and population peaks at $\kna \sim 15 \ent$ and $\kna \sim 150
   \ent$. The effects of PSF smearing and angular resolution on
   best-fit \kna\ values are investigated using mock
   \chandra\ observations and degraded entropy profiles,
   respectively. We find that neither of these effects is sufficient
   to explain the entropy-profile flattening we measure at small
   radii. The influence of profile curvature and number of radial bins
   on best-fit \kna\ is also considered, and we find no indication
   \kna\ is significantly impacted by either. For completeness, we
   include previously unpublished optical spectroscopy of \halpha\ and
   $[N~II]$ emission lines discussed in \citet{haradent}. All data and
   results associated with this work are publicly available via the
   project web site.
\end{abstract}


\section{Introduction}
\label{sec:intro}

The general process of galaxy cluster formation through hierarchical
merging is well understood, but many details, such as the impact of
feedback sources on the cluster environment and radiative cooling in
the cluster core, are not. The nature of feedback operating within
clusters is of great interest because of the implications regarding
the formation of massive galaxies and for the cluster mass-observable
scaling relations used in cosmological studies. Early models of
structure formation which included only gravitation predicted
self-similarity among the galaxy cluster population. These
self-similar models made specific predictions for how the physical
properties of galaxy clusters, such as temperature and luminosity,
should scale with cluster redshift and mass \citep{kaiser86, kaiser91,
  1991ApJ...383...95E, nfw1, nfw2, 1996ApJ...469..494E,
  1997MNRAS.292..289E, 1997ApJ...480...36T, 1998ApJ...503..569E,
  1998ApJ...495...80B}. However, numerous observational studies have
shown clusters do not follow the tight mass-observable scaling
relations predicted by simulations \citep{edge91, 1998MNRAS.297L..57A,
  1998ApJ...504...27M, 1999MNRAS.305..631A, 1999ApJ...520...78H,
  2000ApJ...536...73N, 2001A&A...368..749F}. To reconcile observation
with theory, it was realized non-gravitational effects, such as
heating and radiative cooling in cluster cores, could not be neglected
if models were to accurately replicate the process of cluster
formation \citep[\eg][]{kaiser91, 1991ApJ...383...95E,
  2000ApJ...532...17L, voitbryan, 2002MNRAS.336..409B}.

As a consequence of radiative cooling, best-fit total cluster
temperature decreases while total cluster luminosity increases. In
addition, feedback sources such as active galactic nuclei (AGN) and
galactic winds can drive cluster cores (where most of the cluster flux
originates) away from hydrostatic equilibrium. Thus, at a given mass
scale, radiative cooling and feedback conspire to create dispersion in
otherwise theoretically tight mass-observable correlations like
mass-luminosity and mass-temperature. While considerable progress has
been made both observationally and theoretically in the areas of
understanding, quantifying, and reducing scatter in cluster scaling
relations \citep{1996ApJ...458...27B, 2005ApJ...624..606J, kravtsov06,
2006ApJ...639...64O, nagai07, VV08}, it is still important to
understand how non-gravitational processes, taken as a whole, affect
cluster formation and evolution.

A related issue to the departure of clusters from self-similarity is
that of cooling flows in cluster cores. The core cooling time in
50\%-66\% of clusters is much shorter than both the Hubble time and
cluster age \citep{1984ApJ...285....1S, 1992MNRAS.258..177E, white97,
  1998MNRAS.298..416P, 2005MNRAS.359.1481B}. For such clusters (and
without compensatory heating), radiative cooling will result in the
formation of a cooling flow \citep[see][for a
  review]{fabiancfreview}. Early estimates put the mass deposition
rates from cooling flows in the range of $100-1000 \Msol \pyr$
\citep[\eg][]{1984ApJ...276...38J, 1994MNRAS.270L...1E,
  1998MNRAS.298..416P} However, cooling flow mass deposition rates
inferred from soft X-ray spectroscopy were found to be significantly
less than predicted, without much gas reaching temperatures lower than
$T_{virial}/3$ \citep{tamura01, peterson01, peterson03,
  2004A&A...413..415K}. Irrespective of system mass, the expected
massive torrents of cool gas turned out to be more like cooling
trickles.

In addition to the lack of soft X-ray line emission from cooling
flows, prior methodical searches for the end products of cooling flows
(\ie\ in the form of molecular gas and emission line nebulae) revealed
far less mass is locked-up in cooled by-products than expected
\citep{heckman89, mcnamara90, odea94, voit95}. The disconnects between
observation and theory have been termed ``the cooling flow problem''
and raise the question, ``Where has all the cool gas gone?'' The
substantial amount of observational evidence suggests some combination
of energetic feedback sources, such as AGN outbursts and supernovae
explosions, have heated the ICM to selectively remove gas with a short
cooling time and establish quasi-stable thermal balance in the ICM.

Both the breakdown of self-similarity and the cooling flow problem
point toward the need for a better understanding of cluster feedback
and radiative cooling. Recent revisions to models of how clusters form
and evolve by including feedback sources has led to better agreement
between observation and theory \citep{bower06, croton06, saro06,
  bower08}. The current paradigm regarding the cluster feedback
process holds that AGN are the primary heat delivery mechanism and
that an AGN outburst deposits the requisite energy into the ICM to
retard, and in some cases, possibly quench cooling \citep[see][for a
  review]{mcnamrev}. How the feedback loop functions is still the
topic of much debate, but that AGN are interacting with the hot
atmospheres of clusters is no longer in doubt as evidenced by the
prevalence of ICM bubbles \citep[\eg][]{birzan04,dunn08}, the possible
presence of sound waves \citep{2003MNRAS.344L..43F,
  2008MNRAS.390L..93S}, and large-scale shocks associated with AGN
outbursts \citep{2005ApJ...635..894F, ms0735, 2005ApJ...628..629N}.

One robust observable which has proven useful in studying the effect
of non-gravitational processes is ICM entropy. Taken individually, ICM
temperature and density do not fully reveal a cluster's thermal
history. ICM temperature primarily reflects the depth of a cluster
potential well, while the ICM density mostly reflects the capacity of
the well to compress the gas. However, at constant pressure the
density of a gas is determined by its specific entropy. By rewriting
the expression for the adiabatic index -- which can be expressed as $K
\propto P\rho^{-5/3}$ -- using the observables X-ray temperature
($T_X$) and electron density (\nelec), one can define a new quantity,
$K = T_X n_e^{-2/3}$ \citep{1999Natur.397..135P, davies00}. The
quantity $K$ captures the thermal history of a gas because only gains
and losses of heat energy can change $K$. The expression for $K$ using
observable X-ray quantities is commonly referred to as entropy in the
X-ray cluster literature, but in actuality the classic thermodynamic
specific entropy for a monatomic ideal gas is $s = \ln K^{3/2} +
\mathrm{constant}$.

One important property of gas entropy is that convective stability is
approached in the ICM when $dK/dr \geq 0$. Thus, gravitational
potential wells are giant entropy sorting devices: low entropy gas
sinks to the bottom of the potential well, while high entropy gas
buoyantly rises to a radius at which the ambient gas has equal
entropy. If cluster evolution proceeded under the influence of
gravitation only, then the radial entropy distribution of clusters
would exhibit power-law behavior for $r > 0.1 r_{200}$ with a
constant, low entropy core at small radii \citep{vkb05}. Thus,
large-scale departures of the radial entropy distribution from a
power-law can be used to measure the effect processes such as AGN
heating and radiative cooling have on the ICM. Several studies have
previously found that the radial ICM entropy distribution in some
clusters flattens at $< 0.1 r_{virial}$, or that the core entropy has
much larger dispersion than the entropy at larger radii
\citep{1996ApJ...473..692D, 1999Natur.397..135P, davies00, ponman03,
piffaretti05, radioquiet, pratt06, d06, morandi07}. However, these previous
studies used smaller, focused samples, and to expand the utility of
entropy in understanding cluster thermodynamic history and
non-gravitational processes, we have undertaken a much larger study
utilizing the \chandra\ Data Archive.

In this paper we present the data analysis and results from a
\chandra\ archival project in which we studied the ICM entropy
distribution for \numcluster\ galaxy clusters. We have named this
project the ``Archive of \chandra\ Cluster Entropy Profile Tables'' or
\accept\ for short. In contrast to the sample of nine classic cooling
flow clusters studied in \citet[][hereafter D06]{d06}, \accept\ covers
a broader range of luminosities, temperatures, and morphologies,
focusing on more than just cooling flow clusters. One of our primary
objectives for this project was to provide the research community with
an additional resource to study cluster evolution and confront current
and future ICM models with a comprehensive set of entropy profiles.

We have found that the departure of entropy profiles from a power-law
at small radii is a feature of most clusters, and given high enough
angular resolution, possibly all clusters. We also find that the core
entropy distribution of both the full \accept\ collection and the
Highest X-Ray Flux Galaxy Cluster Sample (\hifl, \citealt{hiflugcs1,
  hiflugcs2}) are bimodal. In a separate letter \citep{haradent}, we
presented results that show indicators of feedback like radio sources
assumed to be associated with AGN and \halpha\ emission are strongly
correlated with core entropy.

A key aspect of this project is the dissemination of all data and
results to the public. We have created a searchable, interactive web
site\footnote{\url{http://www.pa.msu.edu/astro/MC2/accept}} which
hosts all of our results. The \accept\ web site will be continually
updated as new \chandra\ cluster and group observations are archived
and analyzed. The web site provides all data tables, plots, spectra,
reduced \chandra\ data products, reduction scripts, and more. Given
the large number of clusters in our sample, we have omitted figures,
and tables showing/listing results for individual clusters from this
paper and have made them available at the \accept\ web site.

The structure of this paper is as follows: In \S\ref{sec:sample} we
outline initial sample selection criteria and information about the
\chandra\ observations selected under these criteria. Data reduction
is discussed in \S\ref{sec:data}. Spectral extraction and analysis are
discussed in \S\ref{sec:temppr}, while our method for deriving
deprojected electron density profiles is outlined in
\S\ref{sec:dene}. A few possible sources of systematics are discussed
in \S\ref{sec:sys}. Results and discussion are presented in
\S\ref{sec:r&d}. A brief summary is given in \S\ref{sec:summary}. For
this work we have assumed a flat \LCDM\ Universe with cosmology
$\OM=0.3$, $\OL=0.7$, and $\Hn=70\km\ps\pMpc$. All quoted
uncertainties are 90\% confidence.

\section{Data Collection}
\label{sec:sample}

Our sample is collected from observations taken with the
\chandra\ X-ray Observatory \citep{chandra} and which are publicly
available in the \chandra\ Data Archive (CDA) as of August 2008. All
data was taken with the ACIS detectors \citep{acis}, which have a
pixel scale of $\sim 0.492\arcs$ with an on-axis point spread function
(PSF) which is smaller than the detectors' pixel size. ACIS has an
energy resolution of $< 100$ eV for $E \la 2$ keV and $< 300$ eV at
all energies. \chandra's unobscured collecting area is $\sim 1145
\cmsq$ with an effective area of $\sim 600 \cmsq$ around the peak
emission energies of a typical galaxy cluster. At launch ACIS-I and
ACIS-S differed by the better soft-energy sensitivity of ACIS-S, but
in-flight degradation of the CCDs has slowly closed the differences
between the two chip arrays.

We retrieved all data from the CDA listed under the CDA Science
Categories ``clusters of galaxies'' or ``active galaxies.'' As of
submission, we have inspected all CDA clusters of galaxies
observations and analyzed 510 of those observations (14.16 Msec). The
Coma and Fornax clusters have been intentionally left out of our
sample because they are very well studied nearby clusters which
require a more intensive analysis than we undertook in this project.

The data available for some clusters limited our ability to derive an
entropy profile. Calculation of ICM entropy requires measurement of
the gas temperature and density structure as a function of radius
(discussed further in \S\ref{sec:data}). To infer temperatures which
were reasonably well constrained ($\Delta (kT_X) \approx \pm 1.0
\keV$) and to measure more than linear temperature gradients, we
imposed the requirements that each cluster temperature profile have at
least three concentric radial annular bins containing a minimum of
2500 source counts each. A post-analysis check showed our minimum
source counts criterion resulted in a mean $\Delta (kT_X) = 0.87$ keV
for the final sample.

In section \ref{sec:hifl} we cull the flux-limited \hifl\ primary
sample \citep{hiflugcs1, hiflugcs2} from our full archival
collection. The groups M49, NGC 507, NGC 4636, NGC 5044, NGC 5813, and
NGC 5846 are part of the \hifl\ primary sample but were not members of
our initial archival sample. In order to take full advantage of the
\hifl\ primary sample, we analyzed observations of these 6
groups. Note, however, that none of these 6 groups are included in the
general discussion of
\accept.

We were unable to analyze some clusters for this study because of
complications other than not meeting our minimum requirements for
analysis. These clusters were: 2PIGG J0311.8-2655, 3C 129, A168, A514,
A753, A1367, A2634, A2670, A2877, A3074, A3128, A3627, AS0463, APMCC
0421, MACS J2243.3-0935, MS J1621.5+2640, RX J1109.7+2145, RX
J1206.6+2811, RX J1423.8+2404, SDSS J198.070267-00.984433, Triangulum
Australis, and Zw5247.

After applying the temperature profile constraints, adding the 6
\hifl\ groups, and removing troublesome observations, the final sample
presented in this paper contains \numobs\ observations of
\numcluster\ clusters with a total exposure time of \expt. The sample
covers the temperature range $kT_X \sim 1-20$ keV, a bolometric
luminosity range of $L_{bol} \sim 10^{42-46} \ergps$, and redshifts of
$z \sim 0.05-0.89$. Table \ref{tab:sample} lists the general
properties for each observation in \accept.

We also report previously unpublished \halpha\ observations taken by
M. Donahue. These observations do not enter into the analysis
performed in this paper but are used in \citet{haradent}. Since this
paper represents the data of the full project, we include them
here. The new $[N~II]/\halpha$ ratios and \halpha\ fluxes are listed
in Table \ref{tab:newha}. The upper-limits listed in Table
\ref{tab:newha} are $3\sigma$ significance. The observations were
taken with either the 5 m Hale Telescope at the Palomar Observatory,
USA, or the Du Pont 2.5 m telescope at the Las Campanas Observatory,
Chile. All observations were made with a $2\arcs$ slit centered on the
brightest cluster galaxy (BCG) using two position angles: one along
the semi-major axis and one along the semi-minor axis of the
galaxy. The red light (555-798 nm) setup on the Hale Double
Spectrograph used a 316 lines/mm grating with a dispersion of 0.31
nm/pixel and an effective resolution of 0.7-0.8 nm. The Du Pont
Modular Spectrograph setup included a 1200 lines/mm grating with a
dispersion of 0.12 nm/pixel and an effective resolution of 0.3 nm. The
statistical and calibration uncertainties for the observations are
both $\sim 10\%$. The statistical uncertainty arises primarily from
uncertainty in the continuum subtraction.

\section{Data Analysis}
\label{sec:data}

Measuring ICM entropy profiles first requires measurement of ICM
temperature and density profiles. As discussed in \citet{xrayband},
the ICM X-ray peak of the point-source cleaned, exposure-corrected
cluster image was used as the cluster center, unless the iteratively
determined X-ray centroid was more than 70 kpc away from the X-ray
peak, in which case the centroid was used as the radial analysis
zero point (see \cite{xrayband} for more details on centroiding
procedure). The radial temperature structure of each cluster was
measured by fitting a single-temperature thermal model to spectra
extracted from concentric annuli centered on the cluster X-ray
center. To derive the gas density profile, we first deprojected an
exposure-corrected, background-subtracted, point source clean surface
brightness profile extracted in the 0.7-2.0\keV\ energy range to
attain a volume emission density. This emission density, along with
spectroscopic information (count rate and normalization in each
annulus), was then used to calculate gas density. The resulting
entropy profiles were then fit with two models: a simple model
consisting of only a radial power law, and a model which is the sum of
a constant core entropy term, \kna, and the radial power law.

In this paper we cover the basics of deriving gas entropy from X-ray
observables, and direct interested readers to D06 for more in-depth
discussion of our data reprocessing and reduction, and
\citet{xrayband} for details regarding determination of each cluster's
center and how the X-ray background was handled. The only difference
between the data reduction presented in this paper and that of D06 and
\citet{xrayband}, is that we have used newer versions of the \chandra\
X-ray Center (CXC) issued data reduction software (\ciao\ 3.4.1 and
calibration files in the \caldb\ 3.4.0).

\subsection{Temperature Profiles}
\label{sec:temppr}

One of the two components needed to derive a gas entropy profile is
the temperature as a function of radius. We therefore constructed
radial temperature profiles for each cluster in our collection. To
reliably constrain a temperature, and allow for the detection of
temperature structure beyond linear gradients, we required each
temperature profile to have a minimum of three annuli containing 2500
counts each. The annuli for each cluster were generated by first
extracting a background-subtracted cumulative counts profile using 1
ACIS detector pixel width annular bins (1 ACIS pixel $\approx
0.492\arcs$) originating from the cluster center and extending to the
detector edge. We truncated temperature profiles at the radius bounded
by the detector edge, or $0.5 r_{180}$, whichever was
smaller. Truncation occurred at $0.5 r_{180}$ as we are most
interested in the radial entropy behavior of cluster core regions ($r
\la 100$ kpc) and $0.5 r_{180}$ is the approximate radius where
temperature profiles begin to decline at larger radii
\citep{2005ApJ...628..655V}.  Additionally, analysis of diffuse gas
temperature structure at large radii, which spectroscopically is
dominated by background, requires a time consuming,
observation-specific analysis of the X-ray background \cite[see][for a
  detailed discussion on this point]{minggroups}.

Cumulative counts profiles were divided into annuli containing at
least 2500 counts. For well-resolved clusters, the number of counts
per annulus was increased to reduce the resulting uncertainty of
$kT_X$ and, for simplicity, to keep the number of annuli less than 50
per cluster. The method we use to derive entropy profiles is most
sensitive to the surface brightness radial bin size and not the
resolution or uncertainties of the temperature profile. Thus, the loss
of resolution in the temperature profile from increasing the number of
counts per bin, and thereby reducing the number of annuli, has an
insignificant effect on the final entropy profiles and best-fit
entropy models.

Background analysis was performed using the blank-sky datasets
provided in the \caldb. Backgrounds were reprocessed and reprojected
to match each observation. Off-axis chips were used to normalize for
variations of the hard-particle background by comparing blank-sky and
observation 9.5-12\keV\ count rates. Following the analysis described
in \citet{2005ApJ...628..655V}, soft residuals were created and fitted
for each observation to account for the spatially-varying soft
Galactic background \citep[see also][]{xrayband}. The best-fit
spectral model for the residual soft component (scaled for sky area)
was included as an additional, fixed background component during
fitting of cluster spectra. Errors associated with the additional soft
background component were determined by refitting cluster spectra
using the $\pm 1\sigma$ temperatures of the soft background
component's best fit model and then adding the associated error in
quadrature to the final error budget.

For each radial annular region, source and background spectra were
extracted from the target cluster and corresponding normalized
blank-sky dataset. Following standard
\ciao\ techniques\footnote{\url{http://cxc.harvard.edu/ciao/guides/esa.html}}
we created weighted response files (WARF) and redistribution matrices
(WRMF) for each cluster using a flux-weighted map (WMAP) across the
entire extraction region. These files quantify the effective area,
quantum efficiency, and imperfect resolution of the
\chandra\ instrumentation as a function of chip position. Each
spectrum was binned to contain a minimum of 25 counts per energy bin.

Spectra were fitted with \xspec\ 11.3.2ag \citep{xspec} using an
absorbed, single-temperature \mekal\ model \citep{mekal1, mekal2} over
the energy range 0.7-7.0 \keV. Neutral hydrogen column densities,
\nhi, were taken from \citet{dickeylockman}. A comparison between the
\nhi\ values of \citet{dickeylockman} and the higher-resolution
Leiden/Argentine/Bonn (LAB) Survey \citep{lab} revealed that the two
surveys agree to within $\pm 20\%$ for 80\% of the clusters in our
sample. For the other 20\% of the sample, using the LAB value, or
allowing \nhi\ to be free, did not result in best-fit temperatures or
metallicities which differ significantly from fits using the
\citet{dickeylockman} values.

The potentially free parameters of the absorbed thermal model are
\nhi, X-ray temperature, metal abundance normalized to solar
\citep[heavy-element ratios taken from][]{ag89}, and a normalization
($\eta$) which is proportional to the integrated emission measure
within the extraction region,
\begin{equation}
\label{eqn:norm}
\eta = \frac{10^{-14}}{4\pi D_A^2(1+z)^2}\int \nelec \np dV,
\end{equation}
where $D_A$ is the angular diameter distance in cm, $z$ is the
dimensionless cluster redshift, \nelec\ and \np\ are the electron and
proton densities, respectively, in units of $\cm^{-3}$, and $V$ is the
volume of the emission region in $\cm^3$. In all spectral fits the
metal abundance in each annulus was a free parameter and \nhi\ was
fixed to the Galactic value. No systematic error was added during
fitting and thus all quoted errors are statistical only. The statistic
used during fitting was $\chi^2$ (\xspec\ statistics package
\textsc{chi}). All uncertainties were calculated using 90\%
confidence.

More than one observation was available in the archive for some
clusters. We utilized the combined exposure time for these clusters by
first extracting independent spectra, WARFs, WRMFs, normalized
background spectra, and soft residuals for each observation. These
independent spectra were then read into \xspec\ simultaneously and fit
with the same spectral model which had all parameters, except
normalization, tied among the spectra.

Spectral deprojection of ICM temperature should result in slightly
lower temperatures in the central bins of only the clusters with
temperature gradients which increase steeply going out from the
cluster center. For those clusters, the end result would be a slight
lowering of the entropy for the central-most bins. In D06 we studied a
sample of nine ``classic'' cooling flow clusters, all of which have
steep temperature gradients ($T(r)_{max}/T(r)_{min} \sim
1.5-3.5$). Our analysis in D06 showed that spectral deprojection did
not result in significant differences between entropy profiles derived
using projected or deprojected temperature profiles. In light of this
result, and the fact that deprojection requires about a factor of 5
more computing resources and time, we opted not to deproject our
spectra for this phase of the project.

\subsection{Deprojected Electron Density Profiles}
\label{sec:dene}

For predominantly free-free emission, emissivity strongly depends on
density and only weakly on temperature, $\epsilon \propto \rho^2
T^{1/2}$. Since ICM temperatures generally exceed 2.0 keV, the flux
measures in the energy range 0.7-2.0 keV, together with a small
correction for any variations in temperature and metallicity, is
therefore a good diagnostic of ICM density. To reconstruct the
relevant gas density as a function of physical radius, we deprojected
the cluster emission from high-resolution surface brightness profiles
and converted to electron density using normalizations and count rates
taken from the spectral analysis.

We extracted surface brightness profiles from the 0.7-2.0 keV energy
range using concentric annular bins of width $5\arcs$ originating from
the cluster center. Surface brightness profiles were corrected with
observation-specific, normalized radial exposure profiles to remove
the effects of vignetting and exposure time fluctuations. Following
the recommendation in the \ciao\ guide for analyzing extended sources,
exposure maps were created using the monoenergetic value associated
with the observed count rate peak. The more sophisticated method of
creating exposure maps using spectral weights calculated for an
incident spectrum with the temperature and metallicity of the observed
cluster was also tested for a series of clusters covering a broad
temperature range. For the narrow energy band we consider, the chip
response is relatively flat and we find no significant differences
between the two methods. For all clusters, the monoenergetic value
used in creating exposure maps was between $0.8-1.7\keV$.

The 0.7-2.0 keV spectroscopic count rate and spectral normalization
were linearly interpolated from the radial temperature profile grid to
match the surface brightness radial grid. Utilizing the deprojection
technique of \citet{kriss83}, the interpolated spectral parameters
were used to convert observed surface brightness to deprojected
electron density. The conversion from best-fit spectroscopic values to
density intrinsically accounts for temperature and metal abundance
variations which affect the gas emissivity in our selected energy
range. Radial electron density written in terms of relevant quantities
is,
\begin{equation}
\nelec(r) = \sqrt{\frac{(\nelec/\np)~4 \pi [D_A(1+z)]^2~C(r)~\eta(r)}{10^{-14}~f(r)}}
\end{equation}
where $\nelec/n_p \approx 1.2$ for a fully ionized solar abundance
plasma, $C(r)$ is the radial emission density derived from eq. A1 in
\citet{kriss83}, $\eta$ is the interpolated spectral normalization
from eq. \ref{eqn:norm}, $D_A$ is the angular diameter distance, $z$
is cluster redshift, and $f(r)$ is the interpolated spectroscopic
count rate. Cosmic dimming of source surface brightness is accounted
for by the $D_A^2 (1+z)^2$ term. This method of deprojection takes
into account temperature and metallicity fluctuations which affect
observed gas emissivity. Errors for the gas density profile were
estimated using 5000 Monte Carlo simulations of the original surface
brightness profile. The \citet{kriss83} deprojection technique assumes
spherical symmetry. However, D06 showed such an assumption has little
effect on the final entropy profiles \citep[see also][for the low
impact of spherical symmetry assumptions for deriving density
profiles]{2003ApJ...598..190D, 2005MNRAS.359.1481B}.

\subsection{$\beta$-model Fits}
\label{sec:beta}

Noisy surface brightness profiles, or profiles with irregularities
such as inversions or extended flat cores, result in unstable,
unphysical quantities when using an ``onion'' deprojection technique
like that of \citet{kriss83}. For cases where deprojection of the
binned data was problematic, we resorted to fitting the surface
brightness profile with a $\beta$-model \citep{betamodel}, which has
the positive attribute of having an analytic deprojection solution. It
is well known that the $\beta$-model does not precisely represent all
the features of the ICM for clusters of high central surface
brightness \citep{2000MNRAS.311..313E, 2002ApJ...579..571L,
  2007ApJ...665..911H}. However, for the profiles which required a
fit, the $\beta$-model was actually a suitable approximation. These
clusters have low central surface brightness, unlike the classic
cool-core clusters. The single ($N=1$) and double ($N=2$)
$\beta$-models were used in fitting,
\begin{eqnarray}
S_X &=& \displaystyle\sum_{i=1}^N S_i
\left[1+\left(\frac{r}{r_{c,i}}\right)^2\right]^{-3\beta_i+\onehalf}.
\end{eqnarray}
The models were fitted using Craig Markwardt's robust non-linear least
squares minimization IDL
routines\footnote{\url{http://rsinc.com/idl/}}$^{,}$\footnote{\url{http://cow.physics.wisc.edu/~craigm/idl/}}. The
data input to the fitting routines were weighted using the inverse
square of the observational errors. Using this weighting scheme
resulted in reduced \chisq\ values near unity for, on average, the
inner 80\% of the radial range considered. Accuracy of errors output
from the fitting routine were checked against a bootstrap Monte Carlo
analysis of 1000 surface brightness realizations. Both the single- and
double-$\beta$ models were fit to each profile and using the F-test
functionality of
\sherpa\footnote{\url{http://cxc.harvard.edu/ciao3.4/ahelp/ftest.html}}
we determined if the addition of extra model components was justified
given the degrees of freedom and \chisq\ values of each fit. If the
significance was less than 0.05, the extra components were justified
and the double-$\beta$ model was used.

A best-fit $\beta$-model was used in place of the data when deriving
electron density for the clusters listed in Table
\ref{tab:betafits}. These clusters are also flagged in Table
\ref{tab:sample} with the note letter `a.' The best-fit $\beta$-models
and background-subtracted, exposure-corrected surface brightness
profiles are shown in Figure \ref{fig:betamods}. See Appendix
\ref{app:beta} for notes discussing individual clusters. The
disagreement between the best-fit $\beta$-model and the surface
brightness in the central regions for some clusters is also discussed
in Appendix \ref{app:beta}. In short, the discrepancy arises from the
presence of compact X-ray sources, a topic which is addressed in
\S\ref{sec:centsrc}. All clusters requiring a $\beta$-model fit have
core entropy $> 95 \ent$ and the mean best-fit parameters are listed
in Table \ref{tab:bfparams}.

\subsection{Entropy Profiles}
\label{sec:kpr}

Radial entropy profiles were calculated using the widely adopted
formulation $K(r) = kT_x(r)\nelec(r)^{-2/3}$. To create the radial
entropy profiles, the temperature and density profiles must be on the
same radial grid. This was accomplished by interpolating the
temperature profile across the higher-resolution radial grid of the
deprojected electron density profile using IDL's native linear
interpolation routine {\it{interpol}}. Because the density profiles
have higher radial resolution, the central bin of a cluster
temperature profile will span several of the innermost bins of the
density profile. Since we are most interested in the behavior of the
entropy profiles in the central regions, how the interpolation was
performed for the inner regions is important. Thus, temperature
interpolation over the region of the density profile where a single
central temperature bin encompasses several density profile bins was
applied in two ways: (1) as a linear gradient consistent with the
slope of the temperature profile at radii larger than the central
$T_X$ bin ($\Delta T_{center} \ne 0$; `extr' in Table
\ref{tab:kfits}), and (2) as a constant ($\Delta T_{center}=0$; `flat'
in Table \ref{tab:kfits}). Shown in Figure \ref{fig:kcomp} is the
ratio of best-fit core entropy, \kna, using the above two methods. The
five points lying below the line of equality are clusters which are
best-fit by a power-law or have \kna\ statistically consistent with
zero. It is worth noting that both schemes yield statistically
consistent values for \kna\ except for the clusters marked by black
squares which have a ratio significantly different from unity.

The clusters for which the two methods give \kna\ values that
significantly differ all have steep temperature gradients with the
maximum and minimum radial temperatures differing by a factor of
1.3-5.0. Extrapolation of a steep temperature gradient as $r
\rightarrow 0$ results in very low central temperatures (typically
$T_X \leq T_{virial}/3$) which are inconsistent with observations,
most notably \citet{peterson03}. Most important however, is that the
flattening of entropy we observe in the cores of our sample (discussed
in \S\ref{sec:nonzerok0}) is {\bfseries\em{not}} a result of the
method chosen for interpolating the temperature profile. For this
paper, we therefore focus on the entropy results derived assuming a
constant temperature for the central density bins covered by a single
temperature bin.

Uncertainty in $K(r)$ arising from using a single-component
temperature model for each annulus during spectral analysis
contributes negligibly to our final fits and is discussed in detail in
the Appendix of D06. Briefly summarizing D06: the entropy values we
measure at each radius are dominated by the most X-ray luminous
component, which is generally the lowest entropy gas at that
radius. For the best-fit entropy values to be significantly changed,
the volume filling fraction of a higher-entropy component must be
non-trivial ($> 50\%$). As discussed in D06, our results are not
strongly affected by the presence of multiple, low-luminosity gas
phases and are mostly insensitive to X-ray surface brightness
decrements, such as X-ray cavities and bubbles, although in extreme
cases their influence on an entropy profile can be detected (for an
example, see the cluster A2052, also analyzed in D06).

Each entropy profile was fit with two models: a simple model which is
a power-law at large radii and approaches a constant value at small
radii (eq. \ref{eqn:k0}), and a model which is a power-law only
(eq. \ref{eqn:plaw}):
\begin{eqnarray}
K(r) &=& \kna + \khun\ \left(\frac{r}{100 \kpc}\right)^{\alpha}\label{eqn:k0}\\
K(r) &=& \khun\ \left(\frac{r}{100 \kpc}\right)^{\alpha}\label{eqn:plaw}.
\end{eqnarray}
In our entropy models, \kna\ is what we call core entropy, \khun\ is a
normalization for entropy at 100 kpc, and $\alpha$ is the power-law
index. Later in this paper, and in \citet{haradent}, we focus much of
our discussion on the parameter \kna\ so it is worth clarifying what
\kna\ does not represent. \kna\ is not intended to represent the
minimum core entropy or the entropy at $r=0$. Nor does \kna\ capture
the gas entropy which would be measured immediately around an AGN or
in a compact but extended BCG X-ray corona. Instead, \kna\ represents
the typical excess of core entropy above the best fitting power-law at
larger radii. The intentionally simplistic characterization of cluster
core entropy via \kna\ was implemented to make comparing a large
sample of cluster cores less ambiguous. The entropy models were fitted
to the data using Craig Markwardt's IDL routines in the package
MPFIT. The output best-fit parameters and associated errors were
checked using a bootstrap Monte Carlo analysis of 5000 entropy
profile realizations.

The radial range of fitting was truncated at a maximum radius
(determined by eye) to avoid the influence of noisy bins and profile
turnover at large radii which result from instability of our
deprojection method. All the best-fit parameters for each cluster are
listed in Table \ref{tab:kfits}. The mean best-fit parameters for the
full \accept\ sample are given in Table \ref{tab:bfparams}. Also given
in Table \ref{tab:bfparams} are the mean best-fit parameters for
clusters below and above $\kna = 50 \ent$. We show in
\S\ref{sec:bimod} that the cut at $\kna=50 \ent$ is not completely
arbitrary as it approximately demarcates the division between two
distinct populations in the \kna\ distribution.

Some clusters have a surface brightness profile which is comparable to
a double $\beta$-model. Our models for the behavior of $K(r)$ are
intentionally simplistic and are not intended to fully describe all
the features of $K(r)$. Thus, for the small number of clusters with
discernible double-$\beta$ behavior, fitting of the entropy profiles
was restricted to the innermost of the two $\beta$-like
features. These clusters have been flagged in Table \ref{tab:sample}
with the note letter `b.' The best-fit power-law index is typically
much steeper for these clusters, but the outer regions, which we do
not discuss here, have power-law indices which are typical of the rest
of the sample, \ie\ $\alpha \sim 1.2$.

\subsection{Exclusion of Central Sources}
\label{sec:centsrc}

For many clusters in our sample the ICM X-ray peak, ICM X-ray
centroid, BCG optical emission, and BCG infrared emission are
coincident or well within 70 kpc of one another. This made
identification of the cluster center unambiguous in those
cases. However, in some clusters, there is an X-ray point source or
compact X-ray source ($r \la 5$ kpc) found very near ($r < 10$ kpc)
the cluster center and always associated with a galaxy. We identified
\centsrcnum\ clusters with central sources and have flagged them in
Table \ref{tab:sample} with the note letter `d' for AGN and `e' for
compact but resolved sources. The mean best-fit parameters for these
clusters are given in Table \ref{tab:bfparams} under the sample name
`CSE' for ``central source excluded.'' These clusters cover the
redshift range $z = 0.0044-0.4641$ with mean $z = 0.1196 \pm 0.1234$,
and temperature range $kT_X = 1-12$ keV with mean $kT_X = 4.43 \pm
2.53$ keV. For some objects -- such as 3C 295, A2052, A426, Cygnus A,
Hydra A, or M87 -- the source is an AGN and there was no question the
source must be removed.

However, determining how to handle the compact X-ray sources was not
so straightforward. These compact sources are larger than the PSF,
fainter than an AGN, but typically have significantly higher surface
brightness than the surrounding ICM such that the compact source's
extent was distinguishable from the ICM. These sources are most
prominent, and thus the most troublesome, in non-cool core clusters
(\ie\ clusters which are approximately isothermal). They are
troublesome because the compact source is typically much cooler and
denser than the surrounding ICM and hence has an entropy much lower
than the ambient ICM. We believe most of these compact sources to be
X-ray coronae associated with the BCG \citep[see][for discussion of
  BCG coronae]{coronae}.

Without removing the compact sources, we measured radial entropy
profiles and found, for all cases, that $K(r)$ abruptly changes at the
outer edge of the compact source. Including the compact sources in the
measurement of $K(r)$ results in the central cluster region(s)
appearing overdense, and at a given temperature the region will have a
much lower entropy than if the source were excluded. Such a
discontinuity in $K(r)$ results in our simple models of $K(r)$ not
being a good description of the profiles. Aside from producing poor
fits, a significantly lower entropy influences the value of best-fit
parameters because the shape of $K(r)$ is drastically
changed. Obviously, two solutions are available: exclude or keep the
compact sources during analysis.  Deciding what to do with these
sources depends upon what cluster properties we are specifically
interested in quantifying.

The compact X-ray sources discussed in this section are not
representative of the cluster's core entropy; these sources are
representative of the entropy within and immediately surrounding
peculiar BCGs. Our focus for the \accept\ project was to quantify the
entropy structure of the cluster core region and surrounding ICM, not
to determine the minimum entropy of cluster cores or to quantify the
entropy of peculiar core objects such as BCG coronae. Thus, we opted
to exclude these compact sources during our analysis. For a few
extraordinary sources, it was simpler to ignore the central bin of the
surface brightness profile during analysis because of imperfect
exclusion of a compact source's extended emission. These clusters have
been flagged in Table \ref{tab:sample} with the note letter `f.'

It is worth noting that when any source is excluded from the data, the
empty pixels where the source once was were not included in the
calculation of the surface brightness (counts and pixels are both
excluded). Thus, the decrease in surface brightness of a bin where a
source has been removed is not a result of the count to area ratio
being artificially reduced.

\section{Systematics}
\label{sec:sys}

Our models for $K(r)$ were designed so that the best-fit \kna\ values
are a good measure of the entropy profile flattening at small
radii. This flattening could potentially be altered through the
effects of systematics such as PSF smearing and binning of the surface
brightness profile. To quantify the extent to which our
\kna\ values are being affected by these systematics, we have analyzed
mock \chandra\ observations created using the ray-tracing program
MARX\footnote{\url{http://space.mit.edu/CXC/MARX/}}, and also by
analyzing degraded entropy profiles generated from artificially
redshifting well-resolved clusters. In the analysis below, we show
that the lack of clusters with $\kna \la 10 \ent$ at $z \ga 0.1$ is
attributable to resolution effects, but that deviation of an entropy
profile from a power-law, even if only in the central-most bin, cannot
be accounted for by PSF effects. We also discuss the number of
profiles which are reasonably well-represented by the power-law only
profile, and establish that no more than $\sim 10\%$ of the entropy
profiles in \accept\ are consistent with a power-law.

\subsection{PSF Effects}
\label{sec:psf}

To assess the effect of PSF smearing on our entropy profiles, we have
updated the analysis presented in \S4.1 of D06 to use MARX
simulations. In the D06 analysis, we assumed the density and
temperature structure of the cluster core obeyed power-laws with $n_e
\propto r^{-1}$ and $T_X \propto r^{1/3}$. This results in a power-law
entropy profile with $K \propto r$. Further assuming the main emission
mechanism is thermal bremsstrahlung, \ie\ $\epsilon_X \propto
T_X^{1/2}$, yields a surface brightness profile which has the form
$S_X \propto r^{-5/6}$. A source image consistent with these
parameters was created in \idl\ and then input to MARX to create the
mock \chandra\ observations.

The MARX simulations were performed using the spectrum of a 4.0 keV,
$0.3 \approx \Zsol$ abundance \mekal\ model. We have tested using
input spectra with $kT_X = 2-10$ keV with varying abundances and find
the effect of temperature and metallicity on the distribution of
photons in MARX to be insignificant for our discussion here. We have
neglected the X-ray background in this analysis as it is overwhelmed
by cluster emission in the core and is only important at large
radii. Observations for both ACIS-S and ACIS-I instruments were
simulated using an exposure time of 40 ksec. A surface brightness
profile was then extracted from the mock observations using the same
$5\arcs$ bins used on the real data.

For $5\arcs$ bins, we find the difference between the central bins of
the input surface brightness and the output MARX observations to be
less than the statistical uncertainty. One should expect this result,
as the on-axis \chandra\ PSF is $\la 1\arcs$ and the surface
brightness bins we have used on the data are five times this
size. What is most interesting and important though, is that our
analysis using MARX suggests any deviation of the surface brightness
-- and consequently the entropy profile -- from a power-law, even if
only in the central bin, is real and cannot be attributed to PSF
effects. Even for the most poorly resolved clusters, the deviation
away from a power-law we observe in a large majority of our entropy
profiles is not a result of our deprojection technique or systematics.

\subsection{Angular Resolution Effects}
\label{sec:angres}

Another possible limitation in measuring \kna\ is the effect of using
discrete, fixed angular size bins when extracting surface brightness
profiles. This choice may introduce a redshift-dependence into the
best-fit \kna\ values because as redshift increases, a fixed angular
size encompasses a larger physical volume and the value of \kna\ may
increase if the bin includes a broad range of gas entropy. Shown in
Figure \ref{fig:k0res} is a plot of the best-fit \kna\ values for our
entire sample versus redshift.

In the full archival sample, we have a few nearby objects ($z < 0.02$)
with $\kna < 10 \ent$ (numbered in Fig. \ref{fig:k0res}) and only one
at higher redshift -- A1991 ($\kna = 1.53 \pm 0.32$, $z = 0.0587$),
which is a very peculiar cluster \citep{2004ApJ...613..180S}. These
low-$z$, low-\kna\ group-scale objects have been included in our
archival sample because they are well-known. Ignoring those systems,
one can see from Fig. \ref{fig:k0res} that out to $z \approx 0.5$
clusters with $\kna \geq 10 \ent$ are found at all redshifts. The
completeness down to $\kna \approx 10 \ent$ at most redshifts combined
with the low-\kna\ nearby systems raises the question: could the lack
of clusters with $\kna \la 10 \ent$ at $z > 0.02$ be plausibly
explained by resolution effects?

To investigate this question we tested the effect redshift has on our
measurements of \kna\ by culling out the subsample of objects with
$\kna \leq 10 \ent$ and $z \leq 0.1$ and degrading their surface
brightness profiles to mimic the effect of increasing the cluster
redshift. Our test is best illustrated using an example: consider a
cluster at $z = 0.1$. For this cluster, $5\arcs \approx 9$ kpc. Were
the cluster at $z = 0.2$, $5\arcs$ would be $\approx 17$ kpc. To mimic
moving this example cluster from $z = 0.1 \rightarrow 0.2$, we can
extract a new surface brightness profile using a bin size of 17 kpc
instead of $5\arcs$. This procedure will result in a new surface
brightness profile which has the angular resolution for a cluster at a
higher redshift, and subsequent analysis of the entropy profile should
yield information about how redshift affects the best-fit \kna. The
preceding method was used to degrade the profiles of the $\kna \leq 10
\ent$ and $z \leq 0.1$ subsample objects. New surface-brightness bin
sizes were calculated for each cluster over an evenly distributed grid
of redshifts in the range $z = 0.1-0.4$ using step sizes of 0.02.

Our temperature profiles were created using a minimum number of counts
per annulus. Hence, clusters with peaked central surface brightness
will have higher resolution temperature profiles. Thus, in addition to
degrading the surface brightness profiles, the temperature profiles
for each cluster were degraded by starting at the innermost
temperature profile annulus and combining neighboring annuli moving
outward. For each $0.1$ step in our redshift grid the number of annuli
which were combined was increased. For $z=0.1$ two neighboring annuli
were combined, for $z=0.2$ three annuli were combined, for $z=0.3$
four annuli, and five annuli at $z=0.4$. In concordance with our
criterion for creating the original temperature profiles, the number
of annuli in the degraded profiles was not allowed to fall below
3. New spectra were extracted for these enlarged regions and analyzed
following the same procedure detailed in \S\ref{sec:temppr}.

The ensemble of artificially redshifted clusters was analyzed using
the procedure outlined in \S\ref{sec:kpr}. As artificial redshift
increases, the number of radial bins decreases while the size of each
bin increases. Fewer radial bins results in a less detailed sampling
of an entropy profile's overall curvature, while the larger bins mask
the entropy-profile flattening because each bin, particularly the bins
nearest the elbow of an entropy profile, encompass a broad range of
entropy. Over the redshift range $z = 0.1-0.3$, the increased size of
the radial bins (and hence broader range of entropy per bin)
dominates, resulting in entropy profiles which have obvious flattened
cores, but the entropy measured in each bin has
increased. Consequently, best-fit \kna\ also increases, on average, as
$\dkna = 2.12 \pm 1.84$ where \kna\ is the original best-fit value and
$\kna^{\prime}$ is the best-fit value of the degraded profiles. But,
when $z > 0.3$, the degraded entropy profiles severely under sample
both the core flattening and overall profile curvature, resulting in
most entropy profiles resembling power-laws with a centralmost bin
that deviates only slightly from the power-law at larger radii. This
translates into a modest increase of best-fit \kna\ which, on average,
is $\dkna = 0.71 \pm 0.57$. However, there is a caveat to our analysis
of the degraded entropy profiles: the size of the region over which
the original entropy profiles flatten is not uniform. Hence, for
clusters with small flattened cores ($r \la 20$ kpc), degradation of
the profiles will more quickly mask out the flattening, and vice versa
for the clusters with large cores. It is also worth noting that as
redshift increases the best-fit power law indices ($\alpha$) become
shallower (\ie\ significantly less than 1.1), the errors on \kna\ and
$\alpha$ increase, and based on \chisq, the power-law only model fits
drastically improve -- though it is still not a better fit than the
model with \kna.

\subsection{Profile Curvature, Number of Bins, and Exposure Time}
\label{sec:curve}

Our analysis of the degraded entropy profiles suggests that \kna\ is
more sensitive to the value of $K(r)$ in the central bins than it is
to the shape of the profile or the number of radial bins. However, for
completeness we investigate in this section: (1) if there is a
correlation between best-fit \kna\ and the curvature of an entropy
profile, and (2) if the number of radial bins correlates with best-fit
\kna. A systematic correlation of \kna\ with these quantities means
the estimates of \kna\ might be biased by, for example, the curvature
of the temperature profile outside the core or by the signal-to-noise
of an observation.

To check for a possible correlation between best-fit \kna\ and profile
curvature we first calculated average profile curvatures,
$\kappa_A$. For each profile, $\kappa_A$ was calculated using the
standard formulation for the curvature of a function, $\kappa =
\|y^{''}\|/(1+y^{'2})^{3/2}$, where we set $y = K(r) =
\kna+\khun(r/100\kpc)^{\alpha}$. This derivation yields,
\begin{equation}
\kappa_A = \frac{\int\frac{\| 100^{-\alpha} (\alpha-1) \alpha \khun
  r^{\alpha-2}\|}{[1+(100^{-\alpha} \alpha \khun
    r^{\alpha-1})^2]^{3/2}} dr}{\int dr}
\label{eqn:avgcurv}
\end{equation}
where $\alpha$ and \khun\ are the best-fit parameters unique to each
entropy profile. The integral over all space ensures we evaluate the
curvature of each profile in the limit where the profiles have
asymptotically approached a constant at small radii and a power law at
large radii. We find that at any value of \kna\ a large range of
curvatures are covered and that there is no systematic trend in
\kna\ associated with $\kappa_A$ (top left panel of
Fig. \ref{fig:sys}). In addition, plots of best-fit \kna\ versus the
number of bins fit in each entropy profile do not reveal any trends,
only scatter (top right panel of Fig. \ref{fig:sys}).

Our temperature profiles were created using a minimum number of counts
per annulus criterion. One can therefore ask if the length of an
observation or the number of bins in the temperature profile
correlates with best-fit \kna. Shown in the bottom left and right
panels of Fig. \ref{fig:sys} are \kna\ versus the total used exposure
time for that cluster and \kna\ versus number of bins in the
temperature profile, respectively. We do not find trends with \kna\ in
either comparison.

As expected, we do not find any systematic trends with profile shape,
number of bins fit in $K(r)$, exposure time, or number of bins in
$\Tx(r)$ which would significantly affect our best-fit
\kna\ values. Thus, we conclude that the \kna\ values discussed in
this paper are, as intended, an adequate measure of the core entropy,
and that any undetected dependence of \kna\ on profile shape or radial
resolution affect our results at significance levels much smaller than
the measured uncertainties.

\subsection{Power-law Profiles}
\label{sec:quality}

Equation \ref{eqn:k0} is a special case of eq. \ref{eqn:plaw} with
$\kna = 0$, meaning that the models we fit to $K(r)$ are nested. A
comparison between the p-values (shown in Table \ref{tab:kfits}) of
each cluster's best-fit models shows which model exhibits more
agreement with the data. In addition, for each fit in Table
\ref{tab:kfits} we show the deviation in units of sigma,
$\sigma_{\kna}$, of the best-fit \kna\ value from zero. We also show
in Table \ref{tab:bfparams} the number of clusters and the percentage
of the sample which have a \kna\ statistically consistent with zero at
various confidence levels. Table \ref{tab:bfparams} shows that at the
$3\sigma$ significance level $\sim10\%$ of the full \accept\ sample
has a best-fit \kna\ value which is consistent with zero. Moreover,
that there is a systematic trend for a single power-law to be a poor
fit mainly at the smallest radii suggests non-zero \kna\ is not
random.

\section{Results and Discussion}
\label{sec:r&d}

Presented in Figure \ref{fig:splots} is a montage of \accept\ entropy
profiles for different temperature ranges. These figures highlight the
cornerstone result of \accept: a uniformly analyzed collection of
entropy profiles covering a broad range of core entropy. Each profile
is color-coded to represent the global cluster temperature. Plotted in
each panel of Fig. \ref{fig:splots} are the mean profiles representing
$\kna \le 50 \ent$ clusters (dashed-line) and $\kna > 50 \ent$
clusters (dashed-dotted line), in addition to the pure-cooling model
of \citet{voitbryan} (solid black line). The theoretical pure-cooling
curve represents the entropy profile of a 5 keV cluster simulated with
radiative cooling but no feedback and gives us a useful baseline
against which to compare \accept\ profiles.

In the following sections we discuss results gleaned from analysis of
our library of entropy profiles. These results include the departure
of most entropy profiles from a simple radial power-law profile, the
bimodal distribution of core entropy, and the asymptotic convergence
of the entropy profiles to the self-similar $K(r) \propto r^{1.1-1.2}$
power-law at $r \geq 100\kpc$.

\subsection{Non-Zero Core Entropy}
\label{sec:nonzerok0}

Arguably the most striking feature of Figure \ref{fig:splots} is the
departure of most profiles from a simple power-law. Core flattening of
surface brightness profiles (and consequently density profiles) is a
well known feature of clusters (\eg\ \citealt{1984ApJ...276...38J},
\citealt{1999ApJ...517..627M} and \citealt{2000MNRAS.318..715X}). What
is notable in our work however is that, based on comparison of reduced
$\chi^2$ and significance of \kna, very few of the clusters in our
sample have an entropy distribution which is best-fit by the power-law
only model (eq. \ref{eqn:plaw}), rather they are sufficiently
well-described by the model which flattens in the core
(eq. \ref{eqn:k0}).

For clusters with central cooling times shorter than the age of the
cluster, non-zero core entropy is an expected consequence of episodic
heating of the ICM \citep{agnframework}, with AGN as one possible
heating source \citep{1997MNRAS.288..355B, 2000ApJ...532...17L,
2001Natur.414..425V, 2001ApJ...549..832S, 2002MNRAS.332..729C,
2002Natur.418..301B, 2002MNRAS.331..545B, 2002MNRAS.333..145N,
2002ApJ...581..223R, 2002MNRAS.335..610A, 2004MNRAS.348.1105O,
2004ApJ...613..811M, 2004ApJ...615..681R, 2004ApJ...617..896H,
2004MNRAS.355..995D, 2005ApJ...622..847S, pizzolato05,
2006ApJ...643..120B, 2006ApJ...638..659M}. Clusters with cooling times
of order the age of the Universe, however, require other mechanisms to
generate their core entropy, for example via mergers or extremely
energetic AGN outbursts. For the very highest \kna\ values, $\kna >
100 \ent$, the mechanism by which the core entropy came to be so large
is not well understood as it is difficult to boost the entropy of a
gas parcel to $> 100 \ent$ via merger shocks
\citep{2008MNRAS.386.1309M} and would require AGN outburst energies
which have never been observed. We are providing the data and results
of \accept\ to the public with the hope that the research community
finds it a useful new resource to further understand the processes
which result in non-zero cluster core entropy.

\subsection{Bimodality of Core Entropy Distribution}
\label{sec:bimod}

The time required for a gas parcel to radiate away its thermal energy
is a function of the gas entropy. Low entropy gas radiates profusely
and is thus subject to rapid cooling, and vice versa for high entropy
gas. Hence, the distribution of \kna\ is of particular interest
because it is an approximate indicator of the cooling timescale in the
cluster core. The \kna\ distribution is also interesting because it
may be useful in better understanding the physical processes operating
in cluster cores. For example, if processes such as thermal conduction
and AGN feedback are important in establishing the entropy state of
cluster cores, then models which properly incorporate these processes
should approximately reproduce the observed \kna\ distribution.

In the top panel of Figure \ref{fig:k0hist} is plotted the
logarithmically binned distribution of \kna. In the bottom panel of
Figure \ref{fig:k0hist} is plotted the cumulative distribution of
\kna. One can immediately see from these distributions that there are
at least two distinct populations separated by a smaller number of
clusters with $\kna \approx 30-50 \ent$. If the distinct bimodality of
the \kna\ distribution seen in the binned histogram were an artifact
of binning, then the cumulative distribution should be relatively
smooth. But, there is clearly a plateau in the cumulative distribution
which coincides with the division between the two populations at $\kna
\approx 30-50 \ent$. We have tested re-binning the \kna\ histogram
using the optimized binning techniques outlined in \citet{knuthbin}
and \citet{2008arXiv0807.4820H} and find no change in the bimodality
or range of the gap in \kna\ versus using naive fixed-width bins.

To further test for the presence of a bimodal population, we utilized
the KMM test of \citet{kmm1}. The KMM test estimates the probability
that a set of data points is better described by the sum of multiple
Gaussians than by a single Gaussian. We tested the unimodal case
versus the bimodal case under the assumption that the dispersion of
the two Gaussian components are not the same. We have used the updated
KMM code of \citet{kmm2} which incorporates bootstrap resampling to
determine uncertainties for all parameters. A post-analysis comparison
of fits assuming the populations have the same and different
dispersions confirms our initial guess that the dispersions are
different is a better model.

The KMM test, as with any statistical test, is very specific. At
zeroth order, the KMM test simply determines if a population is
unimodal or not, and finds the means of these populations. However,
the dispersions of these populations are subject to the quality of
sampling and the presence of outliers (\eg\ KMM must assign all data
points to a population). The outputs of the KMM test are the best-fit
populations to the data, not necessarily the best-fit populations of
the underlying distribution (hence no goodness of fit is
output). However, the KMM test does output a p-value, and with the
assumption that \chisq\ describes the distribution of the likelihood
ratio statistic, $p$ is the confidence interval for the null
hypothesis.

There are a small number of clusters with $\kna \le 4 \ent$ that when
included in the KMM test significantly change the results. Thus, we
conducted tests including and excluding $\kna \le 4 \ent$ clusters and
provide two sets of best-fit parameters. The results of the bimodal
KMM test neglecting $\kna \le 4 \ent$ clusters were two statistically
distinct peaks at \kmma\ and \kmmb. \kmmc\ clusters were assigned to
the first distribution, while \kmmd\ were assigned to the
second. Including $\kna \le 4 \ent$ clusters, the bimodal KMM test
found populations at \kmmf\ (\kmmh\ clusters) and
\kmmg\ (\kmmi\ clusters). The bimodal KMM test neglecting $\kna \le 4
\ent$ clusters returned \kmme, while the test including all clusters
returned \kmmj. These tiny $p$-values indicate the unimodal
distribution is significantly rejected as the parent distribution of
the observed \kna\ distribution. We also checked for bimodality as a
function of redshift by making cuts in redshift space and running the
KMM test using each new distribution. The KMM test indicated that two
statistically distinct populations were not present when the redshift
range was restricted to clusters with $z > 0.4$. For all other
redshift cuts the \kna\ distribution was bimodal. There are 20
clusters with $z > 0.4$, and we suspected this was too few clusters to
detect two populations. As a test, we randomly selected 20 clusters
from our full sample 1000 times and ran the KMM test. A bimodal
population was found in $~2\%$ of the trials, suggesting the lack of
bimodality at $z > 0.4$ is a result of poor statistics.

We pointed out in \S\ref{sec:kpr} that for some clusters in our
archival sample, the different interpolation schemes for the
centralmost bins of the cluster temperature profiles yielded
significantly different \kna\ values (see Fig. \ref{fig:kcomp}). Using
the \kna\ values derived using temperature profiles which were allowed
to decline in the centralmost bins (see \S\ref{sec:kpr}), we repeated
the above analysis checking for bimodality. We find that bimodality is
present using these \kna\ values and that the best-fit values from the
KMM test are not significantly different for either scheme. Our result
of finding bimodality in the \kna\ population is robust to the choice
of temperature profile interpolation scheme.

One possible explanation for a bimodal core entropy distribution is
that it arises from the effects of episodic AGN feedback and electron
thermal conduction in the cluster core. \citet{agnframework} outlined
a model of AGN feedback whereby outbursts of $\sim 10^{45} \ergps$
occurring every $\sim 10^8 \yrs$ can maintain a quasi-steady core
entropy of $\approx 10-30 \ent$. In addition, very energetic and
infrequent AGN outbursts of $\geq 10^{61} \erg$ can increase the core
entropy into the $\approx 30-50 \ent$ range \citep{agnframework}. This
model of AGN feedback satisfactorily explains the distribution of
$\kna \lesssim 50 \ent$, but depletion of the $\kna = 30-50 \ent$
region and populating $\kna > 50 \ent$ requires more
physics. \citet{conduction} have recently suggested that the dramatic
fall-off of clusters beginning at $\kna \approx 30 \ent$ may be the
result of electron thermal conduction. After \kna\ has exceeded
$\approx 30 \ent$, conduction could severely slow, if not halt, a
cluster's core from appreciably cooling and returning to a core
entropy state with $\kna < 30 \ent$. Merger shocks can then readily
raise \kna\ values to $\ga 100 \ent$. This model is supported by
results presented in \citet{haradent}, \citet{2008ApJ...688..859G},
and \citet{2008ApJ...687..899R} which find that the formation of
thermal instabilities and signatures of ongoing feedback and star
formation are extremely sensitive to the core entropy state of a
cluster.

We acknowledge that \accept\ is not a complete, uniformly selected
sample of clusters. This raises the possibility that our sample is
biased towards clusters that have historically drawn the attention of
observers, such as cooling flows or mergers. If that were the case,
then one reasonable explanation of the \kna\ bimodality is that $\kna
= 30-50 \ent$ clusters have not been the focus of much scientific
interest and thus go unobserved. However, as we show in
\S\ref{sec:hifl}, the complete flux-limited \hifl\ sample is also
bimodal. Nevertheless, flux-limited samples do suffer from some
inadequacies and further study of a carefully selected sample of
clusters, chosen either from our own archival sample or using
representative, rather than complete, samples such as REXCESS
\citep{rexcess}, may be warranted.

\subsection{The \hifl\ Sub-Sample}
\label{sec:hifl}

\accept\ is not a flux-limited or volume-limited sample. To ensure our
results are not affected by an unknown selection bias, we culled the
\hifl\ sample from \accept\ for separate analysis. \hifl\ is a
flux-limited sample ($f_X \ge 2 \times 10^{-11} \flux$) selected from
the {\it{REFLEX}} sample \citep{reflex} with no consideration of
morphology. Thus, at any given luminosity in \hifl\ there is a good
sampling of different morphologies, \ie\ possible bias toward
cool-core clusters or mergers has been removed. The sample also covers
most of the sky with holes near Virgo and the Large and Small
Magellanic Clouds, and has no known incompleteness
\citep{2007A&A...466..805C}. There are a total of 106 objects in
\hifl: 63 in the primary sample and 43 in the extended sample. Of
these 106 objects, no public \chandra\ observations were available for
16 objects (A548e, A548w, A1775, A1800, A3528n, A3530, A3532, A3560,
A3695, A3827, A3888, AS0636, HCG 94, IC 1365, NGC 499, RXCJ
2344.2-0422), 6 objects did not meet our minimum analysis requirements
and were thus insufficient for study (3C 129, A1367, A2634, A2877,
A3627, Triangulum Australis), and as discussed in \S\ref{sec:sample},
Coma and Fornax were intentionally ignored. This left a total of 82
\hifl\ objects which we analyzed, 59 from the primary sample ($\sim
94\%$ complete) and 23 from the extended sample ($\sim 50\%$
complete). The primary sample is the more complete of the two, thus we
focus our following discussion on the primary sample only.

The clusters missing from the primary \hifl\ sample are A1367, A2634,
Coma, and Fornax. The extent to which these 4 clusters can change our
analysis of the \kna\ distribution for \hifl\ is limited.  To alter or
wash-out bimodality, all 4 clusters would need to fall in the range
$\kna = 30-50 \ent$, which is certainly not the case for any of these
clusters. A1367 has been studied by \citet{1998ApJ...500..138D} and
\citet{2002ApJ...576..708S}, with both finding that two sub-clusters
are merging in the cluster. The merger process, and the potential for
associated shock formation, is known to create large increases of gas
entropy \citep{2007MNRAS.376..497M}. Given the combination of low
surface brightness, moderate temperatures ($kT_X = 3.5-5.0$ keV), lack
of a temperature gradient, ongoing merger, and presence of a shock, it
is unlikely A1367 has a core entropy $\la 50 \ent$. A2634 is a very
low surface brightness cluster with the bright radio source 3C 465 at
the center of an X-ray coronae \citep{coronae}. Clusters with
comparable properties to A2634 are not found to have $\kna \la 50
\ent$. Coma and Fornax are known to have core entropy $> 50 \ent$
\citep{2008ApJ...687..899R}.

Shown in Figure \ref{fig:hiflk0} are the log-binned (top panel) and
cumulative (bottom panel) \kna\ distributions of the \hifl\ primary
sample. The bimodality seen in the full \accept\ collection is also
present in the \hifl\ sub-sample. Mean best-fit parameters are given
in Table \ref{tab:bfparams}. We again performed two KMM tests: one
test with, and another test without, clusters having $\kna \le 4
\ent$. For the test including $\kna \le 4 \ent$ clusters we find
populations at \hiflkmma\ (\hiflkmmc\ clusters) and
\hiflkmmb\ (\hiflkmmd\ clusters) with \hiflkmme. Excluding clusters
with $\kna \le 4 \ent$ we find peaks at \hiflkmmf\ and \hiflkmmg, each
having \hiflkmmh\ and \hiflkmmi\ clusters, respectively, and
\hiflkmmj.

\citet{2007hvcg.conf...42H} note a similar core entropy bimodality to
the one we find here. \citet{2007hvcg.conf...42H} discuss two distinct
groupings of objects in a plot of average cluster temperature versus
core entropy, with the dividing point being $K \approx 40 \ent$. Our
results agree with the findings of \citet{2007hvcg.conf...42H}. While
the gaps of \accept\ and \hifl\ do not cover the same \kna\ range, it
is interesting that both gaps appear to be the deepest around $\kna
\approx 30 \ent$. That bimodality is present in both \accept\ and the
unbiased \hifl\ sub-sample suggests bimodality is not the result of
simple archival bias.

\subsection{Distribution of Core Cooling Times}
\label{sec:hifl}

In the X-ray regime, cooling time and entropy are related in that
decreasing gas entropy also means shorter cooling time. Thus, if the
\kna\ distribution is bimodal, the distribution of cooling times
should also be bimodal. We have calculated cooling time profiles from
the spectral analysis using the relation
\begin{equation}
\tcool = \frac{3nkT_X}{2\nelec \nH \Lambda(T,Z)}
\label{eqn:tcool}
\end{equation}
where $n$ is the total number density ($\approx 2.3\nH$ for a fully
ionized plasma), \nelec\ and \nH\ are the electron and proton
densities respectively, $\Lambda(T,Z)$ is the cooling function for a
given temperature and metal abundance, and $3/2$ is a constant
associated with isochoric cooling. The values of the cooling function
for each temperature profile bin were calculated in \xspec\ using the
flux of the best-fit spectral model. Following the procedure discussed
in \S\ref{sec:kpr}, $\Lambda$ and $kT_X$ were interpolated across the
radial grid of the electron density profile. The cooling time profiles
were then fit with a simple model analogous to that used for fitting
$K(r)$:
\begin{equation}
\tcool(r) = t_{c0} + t_{100} \left(\frac{r}{100 \kpc}\right)^{\alpha}
\label{eqn:tc0}
\end{equation}
where $t_{c0}$ is core cooling time and $t_{100}$ is a normalization
at 100 kpc.

The \kna\ distribution can also be used to explore the distribution of
core cooling times. Assuming free-free interactions are the dominant
gas cooling mechanism (\ie\ $\epsilon \propto T^{1/2}$),
\citet{radioquiet} show that entropy is related to cooling time via
the formulation:
\begin{equation}
t_{c0}(\kna) \approx 10^8 \yrs\ \left(\frac{\kna}{10 \keV \cmsq}\right)^{3/2} \left(\frac{kT_X}{5 \keV}\right)^{-1}.
\label{eqn:tck0}
\end{equation}

Shown in Figure \ref{fig:t0} is the logarithmically binned and
cumulative distributions of best-fit core cooling times from
eq. \ref{eqn:tc0} (top panel) and core cooling times calculated using
eq. \ref{eqn:tck0} (bottom panel). The bin widths in both histograms
are 0.20 in log-space. The pile-up of cluster core cooling times below
1 Gyr is well known, for example in \citet{hu85} or more recently in
\citet{dunn08}. In addition, the core cooling times we calculate are
consistent with the results of other cooling time studies, such as
\citet{1998MNRAS.298..416P} or \citet{2008ApJ...687..899R}. However,
what is most important about Fig. \ref{fig:t0} is that the distinct
bimodality of the \kna\ distribution is also present in best-fit core
cooling time, $t_{c0}$. A KMM bimodality test using $t_{c0}$ found
peaks at \tckmma\ and \tckmmb\ with \tckmmc\ and \tckmmd\ objects in
each respective population. The probability that the unimodal
distribution is a better fit was once again exceedingly small,
\tckmme.

The bimodality we observe in the cooling-time distribution is not as
pronounced as what we see in the \kna\ distribution, suggesting that
the bimodality in entropy might be easier to observe. Since cooling
time profiles are more sensitive to the resolution of the temperature
profiles than are the entropy profiles, it may be that resolution
effects more seriously limit the quantification of the true cooling
time of the core. For example, if our temperature interpolation scheme
is too coarse, or averaging over many small-scale temperature
fluctuations significantly increases $t_{c0}$, then $t_{c0}$ would not
be the best approximation of true core cooling time. In which case,
the core cooling times might be shorter and the sharpness and offset
of the distribution gaps may not be as distinct.

\subsection{Slope and Normalization of Power-law Components}
\label{sec:slopes}

Beyond $r \approx 100 \kpc$ the entropy profiles show a striking
similarity in the slope of the power-law component which is
independent of \kna. For the full sample, the mean value of the
power-law normalization at large radii, \alphafs. For clusters with
$\kna < 50 \ent$, the mean \alphaga, and for clusters with $\kna \geq
50 \ent$, the mean \alphagb. Our mean slope of $\alpha \approx 1.2$ is
not statistically different from the theoretical value of $\alpha =
1.1$ found by \citet{tozzi01} using semi-analytic models and $\alpha =
1.2$ found by \citet{vkb05} using models with gravitational effects
only. For the full sample, the mean value of \khunfs. Again
distinguishing between clusters below and above $\kna\ = 50 \ent$, we
find \khunga\ and \khungb, respectively. Scaling each entropy profile
by the cluster virial temperature and virial radius considerably
reduces the dispersion in \khun, but we reserve detailed discussion of
scaling relations for a future paper.

\subsection{Comparison of \accept\ with Other Entropy Studies}
\label{sec:comp}

\subsubsection{Studies Using \xmm}

In \S \ref{sec:angres} we presented our analysis of the angular
resolution effects on entropy profiles. In addition to the analysis
shown there, we have also investigated why previous analyses of
\xmm\ data have found that the entropy profiles of clusters are
adequately fit by simple power laws. For this investigation we have
performed the degradation analysis presented in \S \ref{sec:angres} on
all clusters which have a published entropy profile derived using
\xmm\ data and have been observed with \chandra. These clusters are:
2A 0335+096, A262, A399, A426, A478, A496, A1068, A1413, A1795, A1835,
A1991, A2034, A2052, A2204, A2597, A2717, A3112, A4059, Hydra A,
MKW3S, PKS 0745-191, and Sersic 159-03. \xmm\ analyses of these
clusters were presented in \citet{piffaretti05} and
\citet{pratt06}. Below we briefly highlight some of the important
analysis methods used in these two studies.

\citet{piffaretti05} analyzed \xmm\ data for a sample of 17 cooling
flow clusters in the temperature range $kT_X = 1-7$ keV taken from
\citet{2004A&A...413..415K}. The entropy profiles presented in
\citet{piffaretti05} were derived using the PSF-corrected, deprojected
spectral analysis presented in \citet{2004A&A...413..415K}. The
temperature and density profiles were generated using approximately 8
radial annuli per cluster, in which the spectral analysis was
restricted to the energy range 0.2-10.0 keV. The small number of
annuli used to derive entropy profiles in the \citet{piffaretti05}
analysis results in a much coarser angular scale than is presented in
\accept. \citet{piffaretti05} found no evidence for isentropic cores
in their sample, that the entropy profiles increased monotonically
outward, and that the profiles had a mean power law index of $\alpha =
0.95 \pm 0.02$, which is shallower than the mean $\alpha$ we find in
\accept. However, the width of the innermost radial bin in the
\citet{piffaretti05} analysis was never less than $0.01 r_{virial}$,
and they found the dispersion of entropy in the innermost bins to be
greater than at larger radii, strongly suggesting that profile
flattening in the core was not resolved.

\citet{pratt06} used a sample of 10 relaxed systems observed with
\xmm\ at $z < 0.2$ with temperatures in the range $kT_X \approx 2.5-8$
keV. Entropy profiles were derived using PSF-corrected, deprojected
temperature profiles and gas density profiles calculated from an
analytical model fit to PSF convolved surface brightness profiles
presented in \citet{2005A&A...435....1P}. The parametric models used
in \citet{2005A&A...435....1P} to fit the radial surface-brightness
data were a double $\beta$-model, a $\beta$-model modified to allow
for more centrally concentrated gas densities, and a triple
$\beta$-model with all components having a common $\beta$ value. The
temperature profiles had bin sizes of at least $15\arcs$. Like
\citet{piffaretti05}, \citet{pratt06} found no isentropic cores and
that all the entropy profiles increased monotonically
outward. \citet{pratt06} did however find $< 20\%$ dispersion in
entropy at $r > 0.1r_{200}$ and $> 60\%$ dispersion at $r \sim
0.02r_{200}$ in addition to a mean power law index of $\alpha = 1.08
\pm 0.04$, again suggesting the presence of unresolved flattened
cores. However, \citet{pratt06} do note that, ``the slope of the
[entropy] profile becomes shallower towards the centre in some of the
clusters.'' This suggests that had a power-law model with a core term,
such as \kna, been used, some central flattening might have been
detected. In fact, a few of the \citet{pratt06} entropy profiles, for
example those of A2204 or A2597, clearly lie below the best-fit power
law as they enter the cluster core and then flatten back out in the
central bin, suggesting that they might be better fit with a power-law
plus a constant.

Utilizing the degradation analysis presented in \S \ref{sec:angres},
we repeated that analysis for the subsample of clusters with published
entropy profiles derived from \xmm\ data. We selected the degraded
entropy profiles that had bins sizes similar to the bin sizes used in
previous \xmm\ analyses. For the degraded profiles, we found that core
flattening is harder to detect due to the larger bins. Only clusters
with the largest flattened cores (\eg\ 2A0335, Sersic159, A1413) still
had noticeable entropy-profile curvature, while in contrast, clusters
with the smallest cores (\eg\ A3112, A1991, A4059) were as well fit by
the power-law model as a model with non-zero \kna.

\subsubsection{General Comparison of Results}

There are many published studies of ICM entropy, and in this section
we compare the general trends we find with the results of a few other
studies. The studies with which we compare our results are:
\begin{enumerate}
\item \citet{davies00}: \rosat\ and \asca\ data for 20 bound galaxy
  systems in the redshift range $z \approx 0.08-0.2$ and temperature
  range $kT_X \approx 0.5-14$ keV was used in this
  study. \citet{davies00} clearly show flattened entropy profiles for
  clusters with $K(r) > 100 \ent$ at $r \approx 0.01 r_{virial}$,
  while below this limit they find the entropy profiles trend downward
  like power laws. As we showed in \S\ref{sec:angres} using degraded
  \xmm\ data, the finding of power-law entropy profile behavior at
  small radii is most likely the result of not resolving the small
  flattened entropy cores in cool core clusters.
\item \citet{ponman03}: This study used a sample of 66 systems,
  observed with \rosat\ and \asca, in the redshift range $z=
  0.0036-0.208$ and temperature range $kT_X = 0.5-17$ keV and was the
  largest sample with which we compared our results. In general, the
  entropy profiles presented by \citet{ponman03} flatten inside $0.1
  r_{200}$ irrespective of cluster temperature.
\item \citet{morandi07}: Using \chandra\ data, this study examined 24
  galaxy clusters with $kT_X > 6$ keV in the redshift range
  $z=0.14-0.82$. \citet{morandi07} found the power law indices for
  various subsamples to be in the range $\alpha=1-1.18$, and that all
  of the entropy profiles flatten at $r < 0.5r_{2500}$. They also
  found best-fit \kna\ values in the range $20-300 \ent$.
\end{enumerate}

In general, we find good agreement between the properties of our
entropy profiles and the profiles presented in the papers cited above,
specifically that:
\begin{enumerate}
\item Cluster entropy profiles at $r \ga 0.1 r_{virial}$ are well
  described by an entropy distribution which goes as $K(r) \propto
  r^{1.1-1.2}$.
\item The core regions ($r \la 0.1 r_{virial}$) of clusters approach
  isentropic behavior as $r \rightarrow 0$, or in the cases where the
  observations do not resolve the core regions, the dispersion of
  entropy within the core region is considerably larger than the
  dispersion of the entropy at larger ($r \ga 0.1 r_{virial}$) radii.
\item The above two properties are seen in the entropy profiles of
  clusters over a large range of redshifts ($0.05 \la z \la 0.5$),
  temperatures ($0.5 \keV \la kT_X \la 15 \keV$), and luminosities
  ($10^{43-45}$ ergs s$^{-1}$).
\end{enumerate}  

\section{Summary and Conclusions}
\label{sec:summary}

We have presented intracluster medium entropy profiles for a sample of
\numcluster\ galaxy clusters (\expt) taken from the \chandra\ Data
Archive. We have named this project \accept\ for ``Archive of Chandra
Cluster Entropy Profile Tables.'' The reduced data products, data
tables, figures, cluster images, and results of our analysis for all
clusters and observations are freely available at the \accept\ web
site: \url{http://www.pa.msu.edu/astro/MC2/accept}. We encourage
observers and theorists to utilize this library of entropy profiles in
their own work.

We created radial temperature profiles using spectra extracted from a
minimum of three concentric annuli containing 2500 counts each and
extending to either the chip edge or $0.5 r_{180}$, whichever was
smaller. We deprojected surface brightness profiles extracted from
$5\arcs$ bins over the energy range 0.7-2.0 keV to obtain the electron
gas density as a function of radius. Entropy profiles were calculated
from the density and temperature profiles as $K(r) =
T(r)n(r)^{-2/3}$. Two models for the entropy distribution were then
fit to each profile: a power-law only model (eq. \ref{eqn:plaw}) and
a power-law which approaches a constant value at small radii
(eq. \ref{eqn:k0}).

We have demonstrated that the entropy profiles for the majority of
\accept\ clusters are well-represented by the model which approaches a
constant entropy, \kna, in the core. The entropy profiles of
\accept\ are also remarkably similar at radii greater than 100 kpc,
and asymptotically approach the self-similar pure-cooling curve ($r
\propto 1.2$) with a slope of \alphafs\ (the dispersion here is in the
sample, not in the uncertainty of the measurement). We also find that
the distribution of \kna\ for the full archival sample is bimodal with
the two populations separated by a poorly populated region between
$\kna \approx 30-50 \ent$. After culling out the primary
\hifl\ sub-sample of \citet{hiflugcs1}, we find the \kna\ distribution
of this complete sub-sample also to be bimodal, indicating that the
bimodality we find in our larger sample does not result from archival
bias.

When we compared our results with those of a few other entropy
studies, specifically \citet{davies00}, \citet{ponman03},
\citet{piffaretti05}, \citet{pratt06}, and \citet{morandi07}, we found
the same general trends, noting however that \citet{piffaretti05} and
\citet{pratt06} did not specifically find isentropic cores. However,
those two studies did find large dispersion of entropy in the core
region ($r < 0.1 r_{virial}$), suggesting that the broader bins used
for analyzing the \xmm\ data resulted in flattened entropy profiles
not being resolved like they are using finer radial resolution and
\chandra\ data.

Two core cooling times were derived for each cluster: (1) cooling time
profiles were calculated using eq. \ref{eqn:tcool} and each cooling
time profile was then fit with eq. \ref{eqn:tc0} returning a best-fit
core cooling time, $t_{c0}$; (2) Using best-fit \kna\ values, entropy
was converted to a core cooling time, $t_{c0}(\kna)$ using
eq. \ref{eqn:tck0}. We find the distributions of both core cooling
times to be bimodal. Comparison of the core cooling times from method
(1) and (2) reveals that the gap in the bimodal cooling time
distributions occur over different timescales, $\sim 2-3$ Gyrs for
$t_{c0}$, and $\sim 0.7-1$ for $t_{c0}(\kna)$, but this offset may be
the result of resolution limitations.

After analyzing an ensemble of artificially redshifted entropy
profiles, we find the lack of $\kna \la 10 \ent$ clusters at $z > 0.1$
is most likely a result of resolution effects. Investigation of
possible systematics affecting best-fit \kna\ values, such as profile
curvature and number of profile bins, revealed no trends which would
significantly affect our results. We came to the conclusion that
\kna\ is an acceptable measure of average core entropy and is not
overly influenced by profile shape or radial resolution. We also find
that $\sim90\%$ of the sample clusters have a best-fit \kna\ more than
$3\sigma$ away from zero.

Our results regarding non-zero core entropy and \kna\ bimodality
support the sharpening picture of how feedback and radiative cooling
in clusters alter global cluster properties and affect massive galaxy
formation. Among the many models of AGN feedback, \citet{agnframework}
outlined a model which specifically addresses how AGN outbursts
generate and sustain non-zero core entropy in the regime of $\kna \la
30 \ent$ \citep[see also][]{kaiser03}. In addition, if electron
thermal conduction is an important process in clusters, then there
exists a critical entropy threshold below which conduction is no
longer efficient at wiping out thermal instabilities, the consequences
of which should be a bimodal core entropy distribution and a
sensitivity of cooling by-product formation (like star formation and
AGN activity) to this entropy threshold \citep{conduction,
  2008ApJ...688..859G}. We show in \citet{haradent} that indicators of
feedback like \halpha\ and radio emission are extremely sensitive to
the lower bound of the gap in the bimodal distribution at $\kna
\approx 30 \ent$.

Many details are still missing from the emerging picture of the
entropy life cycle in clusters, and there are many open questions
regarding the evolution of the ICM and how thermal instabilities form
in cluster cores. It is still unclear how clusters with very high core
entropy ($\kna > 100 \ent$) are produced. Is an early episode of
preheating necessary? And while resolution has restricted our ability
to investigate a possible evolution of \kna\ with redshift (which
would suggest evolution in the cool-core cluster population), there
may be other observational proxies which tightly correlate with
\kna\ and could then be used to study cluster cores at high-$z$. It is
also becoming clear that the role of ICM magnetic fields can no longer
be ignored. More specifically, how magnetohydrodynamic instabilities,
such as MTI \citep{2000ApJ...534..420B, 2008ApJ...673..758Q} and HBI
\citep{2008ApJ...677L...9P}, might impact the entropy structure of the
ICM and formation of thermal instabilities needs to be investigated
more thoroughly. We hope \accept\ will be a useful resource in
studying these questions.

\acknowledgements

KWC was supported in this work through \chandra\ X-ray Observatory
Archive grants AR-6016X and AR-4017A. MD and MS acknowledge support
from the NASA LTSA program NNG-05GD82G. The \chandra\ X-ray
Observatory Center is operated by the Smithsonian Astrophysical
Observatory for and on behalf of NASA under contract NAS8-03060. KWC
thanks Chris Waters for supplying and supporting his new KMM code, Jim
Linnemann for helpful suggestions regarding the error and statistical
analysis presented in this paper, and Brian McNamara for useful
discussions. We especially thank Gabriel Pratt for sharing entropy
profiles. We also thank the anonymous referee who's comments greatly
improved the content of the paper. This research has made use of
software provided by the Chandra X-ray Center in the application
packages \ciao, \chips, and \sherpa. This research has made use of the
NASA/IPAC Extragalactic Database which is operated by the Jet
Propulsion Laboratory, California Institute of Technology, under
contract with NASA. This research has also made use of NASA's
Astrophysics Data System. Some software was obtained from the High
Energy Astrophysics Science Archive Research Center, provided by
NASA's Goddard Space Flight Center.

{\it Facilities:} \facility{CXO (ACIS)}, \facility{Du Pont (Modular
  Spectrograph)}, \facility{Hale (Double Spectrograph)}

\bibliography{cavagnolo}

\LongTables
\input{appendices.tex}
\clearpage
\input{sample_dataset.tex}
\clearpage
\input{betafits.tex}
\clearpage
\input{newha.tex}
\clearpage
\input{bfparams.tex}
\clearpage
\input{kfit.tex}
\clearpage
\input{figs.tex}

\end{document}

%% file: appendices.tex
\begin{appendix}

\section{Notes on clusters requiring $\beta$-model fit}
\label{app:beta}

\begin{description}
\item[Abell 119 ($z=0.0442$):] This is a highly diffuse cluster
  without a prominent cool core. The large core region and slowly
  varying surface brightness made deprojection highly unstable. We
  have excluded a small source at the very center of the BCG. The
  exclusion region for the source is $\approx 2.2\arcsec$ in radius
  which at the redshift of the cluster is $\sim 2$ kpc. This cluster
  required a double $\beta$-model.

\item[Abell 160 ($z=0.0447$):] The highly asymmetric, low surface
  brightness of this cluster resulted in a noisy surface brightness
  profile that could not be deprojected. This cluster required a
  double $\beta$-model. The BCG hosts a compact X-ray source. The
  exclusion region for the compact source has a radius of $\sim
  5\arcsec$ or $\sim 4.3$ kpc. The BCG for this cluster is not
  coincident with the X-ray centroid and hence is not at the
  zero-point of our radial analysis.

\item[Abell 193 ($z=0.0485$):] This cluster has an azimuthally
  symmetric and a very diffuse ICM centered on a BCG which is
  interacting with a companion galaxy. In Fig. \ref{fig:betamods} one
  can see that the central three bins of this cluster's surface
  brightness profile are highly discrepant from the best-fit
  $\beta$-model. This is a result of the BCG being coincident with a
  bright, compact X-ray source. As we have concluded in
  \ref{sec:centsrc}, compact X-ray sources are excluded from our
  analysis as they are not the focus of our study here. Hence we have
  used the best-fit $\beta$-model in deriving $K(r)$ instead of the
  raw surface brightness.

\item[Abell 400 ($z=0.0240$):] The two ellipticals at the center of
  this cluster have compact X-ray sources which are excluded during
  analysis. The core entropy we derive for this cluster is in
  agreement with that found by \cite{2006A&A...453..433H} which
  supports the accuracy of the $\beta$-model we have used.

\item[Abell 1060 ($z=0.0125$):] There is a distinct compact source
  associated with the BCG in this cluster. The ICM is also very faint
  and uniform in surface brightness making the compact source that
  much more obvious. Deprojection was unstable because of imperfect
  exclusion of the source.

\item[Abell 1240 ($z=0.1590$):] The surface brightness of this cluster
  is well-modeled by a $\beta$-model. There is nothing peculiar worth
  noting about the BCG or the core of this cluster.

\item[Abell 1736 ($z=0.0338$):] Another ``boring'' cluster with a very
  diffuse low surface brightness ICM, no peaky core, and no signs of
  merger activity in the X-ray. The noisy surface brightness profile
  necessitated the use of a double $\beta$-model. The BCG is
  coincident with a very compact X-ray source, but the BCG is offset
  from the X-ray centroid and thus the central bins are not adversely
  affected. The radius of the exclusion region for the compact source
  is $\approx 2.3\arcsec$ or $1.5$ kpc.

\item[Abell 2125 ($z=0.2465$):] Although the ICM of this cluster is
  very similar to the other clusters listed here (\ie\ diffuse, large
  cores), A2125 is one of the more compact clusters. The presence of
  several merging sub-clusters \citep{1997ApJ...487L..13W,
    2004ApJ...611..821W} to the NW of the main cluster form a diffuse
  mass which cannot rightly be excluded. This complication yields
  inversions of the deprojected surface brightness profile if a double
  $\beta$-model is not used.

\item[Abell 2255 ($z=0.0805$):] This is a very well studied merger
  cluster \citep{1995ApJ...446..583B, 1997A&A...317..432F}. The core
  of this cluster is very large ($r > 200$ kpc). Such large extended
  cores cannot be deprojected using our methods because if too many
  neighboring bins have approximately the same surface brightness,
  deprojection results in bins with negative or zero value. The
  surface brightness for this cluster is well modeled as a $\beta$
  function.

\item[Abell 2319 ($z=0.0562$):] A2319 is another well studied merger
  cluster \citep{1997NewA....2..501F, 1999ApJ...525L..73M} with a very
  large core region ($r > 100$ kpc) and a prominent cold front
  \citep{2004ApJ...604..604O}. Once again, the surface brightness
  profile is well-fit by a $\beta$-model.

\item[Abell 2462 ($z=0.0737$):] This cluster is very similar in
  appearance to A193: highly symmetric ICM with a bright, compact
  X-ray source embedded at the center of an extended diffuse ICM. The
  central compact source has been excluded from our analysis with a
  region of radius $\approx 1.5\arcsec$ or $\sim 3$ kpc. The central
  bin of the surface brightness profile is most likely boosted above
  the best-fit double $\beta$-model because of faint extended emission
  from the compact source which cannot be discerned from the ambient
  ICM.

\item[Abell 2631 ($z=0.2779$):] The surface brightness profile for
  this cluster is rather regular, but because the cluster has a large
  core it suffers from the same unstable deprojection as A2255 and
  A2319. The ICM is symmetric about the BCG and is incredibly uniform
  in the core region. We did not detect or exclude a source at the
  center of this cluster, but under heavy binning the cluster image
  appears to have a source coincident with the BCG, and the slightly
  higher flux in central bin of the surface brightness profile may be
  a result of an unresolved source.

\item[Abell 3376 ($z=0.0456$):] The large core of this cluster ($r >
  120$ kpc) makes deprojection unstable and a $\beta$-model must be
  used.

\item[Abell 3391 ($z=0.0560$):] The BCG is coincident with a compact
  X-ray source. The source is excluded using a region with radius
  $\approx 2\arcsec$ or $\sim 2$ kpc. The large uniform core region
  made deprojection unstable and thus required a $\beta$-model fit.

\item[Abell 3395 ($z=0.0510$):] The surface brightness profile for
  this cluster is noisy resulting in deprojection inversions and
  requiring a $\beta$-model fit. The BCG of this cluster has a compact
  X-ray source and this source was excluded using a region with radius
  $\approx 1.9\arcsec$ or $\sim 2$ kpc.

\item[MKW 08 ($z=0.0270$):] MKW 08 is a nearby large group/poor
  cluster with a pair of interacting elliptical galaxies in the
  core. The BCG falls directly in the middle of the ACIS-I detector
  gap. However, despite the lack of proper exposure, CCD dithering
  reveals that a very bright X-ray source is associated with the
  BCG. A double $\beta$-model was necessary for this cluster because
  the low surface brightness of the ICM is noisy and deprojection is
  unstable.

\item[RBS 461 ($z=0.0290$):] This is another nearby large group/poor
  cluster with an extended, diffuse, axisymmetric, featureless ICM
  centered on the BCG. The BCG is coincident with a compact source
  with size $r \approx 1.7$ kpc. This source was excluded during
  reduction. The $\beta$-model is a good fit to the surface brightness
  profile.
\end{description}

\end{appendix}

%% file: sample_dataset.tex
\begin{deluxetable}{lcccccccc}
\tablewidth{0pt}
\tabletypesize{\scriptsize}
\tablecaption{Summary of Sample\label{tab:sample}}
\tablehead{\colhead{Cluster} & \colhead{Obs. ID} & \colhead{R.A.} & \colhead{Decl.} & \colhead{Exposure Time} & \colhead{ACIS} & \colhead{$z$} & \colhead{$kT_{X}$} & \colhead{Notes}\\
\colhead{} & \colhead{} & \colhead{hr:min:sec} & \colhead{$\mydeg:\arcm:\arcs$} & \colhead{ksec} & \colhead{} & \colhead{} & \colhead{keV} & \colhead{}\\
\colhead{{(1)}} & \colhead{{(2)}} & \colhead{{(3)}} & \colhead{{(4)}} & \colhead{{(5)}} & \colhead{{(6)}} & \colhead{{(7)}} & \colhead{{(8)}} & \colhead{{(9)}}}
\startdata
 \object{1E0657 56} & \dataset[ADS/Sa.CXO#obs/3184]{3184} & 06:58:29.627 & -55:56:39.79 & 87.5 & I3 & 0.2960 & 11.64 & \nodata\\
 & \dataset[ADS/Sa.CXO#obs/5356]{5356} & \nodata & \nodata & 97.2 & I2 & \nodata & \nodata & \nodata\\
 & \dataset[ADS/Sa.CXO#obs/5361]{5361} & \nodata & \nodata & 82.6 & I3 & \nodata & \nodata & \nodata\\
 \object{2A 335+096} & \dataset[ADS/Sa.CXO#obs/919]{919} & 03:38:41.105 & +09:58:00.66 & 19.7 & S3 & 0.0347 & 2.88 & \nodata\\
 \object{2PIGG J0011.5-2850} & \dataset[ADS/Sa.CXO#obs/5797]{5797} & 00:11:21.623 & -28:51:14.44 & 19.9 & I3 & 0.0753 & 5.15 &      f\\
 \object{2PIGG J2227.0-3041} & \dataset[ADS/Sa.CXO#obs/5798]{5798} & 22:27:54.560 & -30:34:34.84 & 22.3 & I2 & 0.0729 & 2.79 & \nodata\\
 \object{3C 28.0} & \dataset[ADS/Sa.CXO#obs/3233]{3233} & 00:55:50.401 & +26:24:36.47 & 49.7 & I3 & 0.1952 & 5.53 & \nodata\\
 \object{3C 295} & \dataset[ADS/Sa.CXO#obs/2254]{2254} & 14:11:20.280 & +52:12:10.55 & 90.9 & I3 & 0.4641 & 5.16 &      d\\
\enddata
\tablecomments{Col. (1) Cluster name; Col. (2) CXC CDA Observation Identification Number; Col. (3) R.A. of cluster center; Col. (4) Decl. of cluster center; Col. (5) exposure time; Col. (6) CCD location of cluster center; Col. (7) redshift; Col. (8) average cluster temperature; Col. (9) best-fit core entropy measured in this work; Col. (10) assigned notes: `a' Clusters analyzed using the best-fit $\beta$-model for the surface brightness profiles (discussed in \S\ref{sec:dene}); `b' Clusters with complex surface brightness of which only the central regions were used in fitting $K(r)$; `c' Clusters only used during analysis of the \hifl\ sub-sample (discussed in \S\ref{sec:hifl}); `d' Clusters with central AGN removed during analysis (discussed in \S\ref{sec:centsrc}); `e' Clusters with central compact source removed during analysis (discussed in \S\ref{sec:centsrc}); `f' Clusters with central bin ignored during fitting (discussed in \S\ref{sec:centsrc})..}
\end{deluxetable}

%% file: betafits.tex
\begin{deluxetable}{lcccccccc}
\tablewidth{0pt}
\tabletypesize{\scriptsize}
\tablecaption{Summary of $\beta$-Model Fits\label{tab:betafits}}
\tablehead{\colhead{Cluster} & \colhead{$S_{01}$} & \colhead{$r_{c1}$} & \colhead{$\beta_{1}$} & \colhead{$S_{02}$} & \colhead{$r_{c2}$} & \colhead{$\beta_{2}$} & \colhead{D.O.F.} & \colhead{$\chi_{\mathrm{red}}^2$}\\
\colhead{} & \colhead{10$^{-6}$ cts s$^{-1}$ arcsec$^{2}$} & \colhead{\arcsec} & \colhead{} & \colhead{10$^{-6}$ cts s$^{-1}$ arcsec$^{2}$} & \colhead{\arcsec} & \colhead{} & \colhead{} & \colhead{}\\
\colhead{{(1)}} & \colhead{{(2)}} & \colhead{{(3)}} & \colhead{{(4)}} & \colhead{{(5)}} & \colhead{{(6)}} & \colhead{{(7)}} & \colhead{{(8)}} & \colhead{{(9)}}
}
\startdata
Abell 119 &  4.93 $\pm$  0.73 &  39.1 $\pm$  15.3 &  0.34 $\pm$  0.07 &  3.52 $\pm$  0.96 & 735.2 $\pm$ 479.4 &  1.27 $\pm$  1.27 &    52 &  1.76\\
Abell 160 &  2.32 $\pm$  0.27 &  53.4 $\pm$  11.1 &  0.57 $\pm$  0.12 &  1.29 $\pm$  0.22 & 284.0 $\pm$  52.2 &  0.74 $\pm$  0.10 &    90 &  1.18\\
Abell 193 & 24.72 $\pm$  1.62 &  80.8 $\pm$   2.2 &  0.43 $\pm$  0.01 & \nodata & \nodata & \nodata &    38 &  0.43\\
Abell 400 &  4.66 $\pm$  0.09 & 151.3 $\pm$   6.4 &  0.42 $\pm$  0.01 & \nodata & \nodata & \nodata &    96 &  0.57\\
Abell 1060 & 21.95 $\pm$  0.44 &  93.5 $\pm$   8.1 &  0.35 $\pm$  0.01 & \nodata & \nodata & \nodata &    42 &  1.44\\
Abell 1240 &  1.58 $\pm$  0.07 & 247.9 $\pm$  46.9 &  1.01 $\pm$  0.22 & \nodata & \nodata & \nodata &    58 &  1.58\\
Abell 1736 &  3.81 $\pm$  0.56 &  55.6 $\pm$  16.1 &  0.42 $\pm$  0.12 &  2.49 $\pm$  0.47 & 1470.0 $\pm$  87.2 &  5.00 $\pm$  0.73 &    35 &  1.58\\
Abell 2125 &  3.50 $\pm$  0.20 &  26.0 $\pm$   4.9 &  0.49 $\pm$  0.05 &  1.02 $\pm$  0.13 & 159.9 $\pm$   9.2 &  1.32 $\pm$  0.16 &    35 &  0.33\\
\enddata
\tablecomments{Col. (1) Cluster name; col. (2) central surface brightness of first component; col. (3) core radius of first component; col. (4) $\beta$ parameter of first component; col. (5) central surface brightness of second component; col. (6) core radius of second component; col. (7) $\beta$ parameter of second component; col. (8) model degrees of freedom; and col. (9) reduced chi-squared statistic for best-fit model.}
\end{deluxetable}

%% file: newha.tex
\begin{deluxetable}{lcccc}
\tabletypesize{\scriptsize}
\tablecaption{M. Donahue's \halpha\ Observations.\label{tab:newha}}
\tablewidth{0pt}
\tablehead{
  \colhead{Cluster} & \colhead{Telescope} & \colhead{$z$} & \colhead{$[NII]$/\halpha} & \colhead{\halpha\ Flux}\\
  \colhead{}      & \colhead{}        & \colhead{}  & \colhead{}               & \colhead{$10^{-15}$ \flux}
}
\startdata
Abell 85     & PO & 0.0558 & 2.67    &    0.581\\
Abell 119    & LC & 0.0442 & \nodata & $<$0.036\\
Abell 133    & LC & 0.0558 & \nodata &    0.88\\
Abell 496    & LC & 0.0328 & 2.50    &    2.90\\
Abell 1644   & LC & 0.0471 & \nodata &    1.00\\
Abell 1650   & LC & 0.0843 & \nodata & $<$0.029\\
Abell 1689   & LC & 0.1843 & \nodata & $<$0.029\\
Abell 1736   & LC & 0.0338 & \nodata & $<$0.026\\
Abell 2597   & PO & 0.0854 & 0.85    &    29.7\\
Abell 3112   & LC & 0.0720 & 2.22    &    2.66\\
Abell 3158   & LC & 0.0586 & \nodata & $<$0.036\\
Abell 3266   & LC & 0.0590 & 1.62    & $<$0.027\\
Abell 4059   & LC & 0.0475 & 3.60    &    2.22\\
Cygnus A     & PO & 0.0561 & 1.85    &    28.4\\
EXO 0422-086 & LC & 0.0397 & \nodata & $<$0.031\\
Hydra A      & LC & 0.0522 & 0.85    &    13.4\\
PKS 0745-191 & LC & 0.1028 & 1.02    &    10.4
\enddata
\tablecomments{The abbreviation ``PO'' denotes observations taken on
  the 5 m Hale Telescope at the Palomar Observatory, USA, while ``LC''
  are observations taken on the DuPont 2.5 m telescope at the Las
  Campanas Observatory, Chile. Upperlimits for \halpha\ fluxes are
  $3\sigma$.}
\end{deluxetable}

%% file: bfparams.tex
\begin{deluxetable}{lcccccccc}
\tabletypesize{\scriptsize}
\tablecaption{Statistics of Best-Fit Parameters\label{tab:bfparams}}
\tablewidth{0pt}
\tablehead{
\colhead{Sample} & \colhead{$N_{obj}$} & \colhead{\kna} & \colhead {$K_{12}$} & \colhead{\khun} & \colhead{$\alpha$} & \multicolumn{3}{c}{$N_{\kna=0}$}\\
\colhead{} & \colhead{} & \colhead{\ent} & \colhead{\ent} & \colhead{\ent} & \colhead{} & \colhead{$1\sigma$} & \colhead{$2\sigma$} & \colhead{$3\sigma$}\\
\colhead{(1)} & \colhead{(2)} & \colhead{(3)} & \colhead{(4)} & \colhead{(5)} & \colhead{(6)} & \colhead{(7)} & \colhead{(8)} & \colhead{(9)}
}
\startdata
\multicolumn{9}{c}{All \kna}\\
\hline
\accept\       & 233 & $72.9 \pm 33.7$ & $91.6 \pm 35.7$ & $126 \pm 45$   & $1.21 \pm 0.39$ & 4 (2\%) & 12 (5\%)  & 24 (11\%)\\
\hifl\         & 59  & $62.3 \pm 32.7$ & $87.2 \pm 34.5$ & $166 \pm 65$   & $1.18 \pm 0.38$ & 1 (2\%) & 3  (5\%)  & 4  (7\%) \\
CSE            & 37  & $61.9 \pm 27.4$ & $81.6 \pm 31.3$ & $132 \pm 45$   & $1.19 \pm 0.39$ & 1 (3\%) & 2  (5\%)  & 6  (16\%)\\
$\beta$ Models & 17  & $220  \pm 74$   & $230  \pm 76.9$ & $67.4\pm 27.0$ & $1.45 \pm 0.47$ & \nodata & \nodata   & \nodata  \\[0.25cm]
\hline
\multicolumn{9}{c}{$4 \ent < \kna \le 50 \ent$}\\
\hline
\accept\       & 99  & $17.5 \pm 5.8$  & $31.2 \pm 10.3$ & $148 \pm 49$   & $1.21 \pm 0.39$ & 3 (3\%) & 5  (5\%)  & 10 (10\%)\\
\hifl\         & 25  & $13.6 \pm 4.6$  & $29.4 \pm 9.63$ & $174 \pm 57$   & $1.15 \pm 0.37$ & 0 (0\%) & 0  (0\%)  & 0  (0\%) \\
\enddata
\tablecomments{Listed here are the mean best-fit parameters of the
  model $K(r) = \kna + \khun (r/100 \kpc)^{\alpha}$ for various
  sub-groups of the full \accept\ sample. Each sub-group is labeled in
  the table. The 'CSE' sample are the clusters with a central source
  excluded (discussed in \S\ref{sec:centsrc}). The $K_{12}$ values
  represent the entropy at 12 kpc and are calculated from the best-fit
  models. Col. (1) Sample being considered; col. (2) number of objects
  in the sub-group; col. (3) mean best-fit \kna; col. (4) mean entropy
  at 12 kpc; col. (5) mean best-fit \khun; and col. (6) mean best-fit
  power-law index; cols. (7,8,9) number of clusters consistent with
  $\kna = 0 \ent$ at $1\sigma$, $2\sigma$, and $3\sigma$ significance,
  respectively. Percentage of the sub-group represented by each is
  also listed.}
\end{deluxetable}

%% file: kfit.tex
\begin{deluxetable}{lcccccccccc}
\tablewidth{0pt}
\tabletypesize{\scriptsize}
\tablecaption{Summary of Entropy Profile Fits\label{tab:kfits}}
\tablehead{\colhead{Cluster} & \colhead{Method} & \colhead{$N_{bins}$} & \colhead{$r_{max}$} & \colhead{\kna} & \colhead{$\sigma_{\kna} > 0$} & \colhead{\khun} & \colhead{$\alpha$} & \colhead{DOF} & \colhead{\chisq} & \colhead{p-value}\\
\colhead{} & \colhead{} & \colhead{} & \colhead{Mpc} & \colhead{\ent} & \colhead{} & \colhead{\ent} & \colhead{} & \colhead{} & \colhead{} & \colhead{}\\
\colhead{{(1)}} & \colhead{{(2)}} & \colhead{{(3)}} & \colhead{{(4)}} & \colhead{{(5)}} & \colhead{{(6)}} & \colhead{{(7)}} & \colhead{{(8)}} & \colhead{{(9)}} & \colhead{{(10)}} & \colhead{{(11)}}
}
\startdata
1E0657 56 &   extr &     48 &   1.00 &  299.4 $\pm$   19.6 &   15.3 &   20.5 $\pm$    7.0 &   1.84 $\pm$   0.16 &     45 &  42.09 & 5.96e-01\\
 &      - & - & - &    0.0 & - &  277.9 $\pm$   14.5 &   0.60 $\pm$   0.04 &     46 & 146.18 & 2.31e-12\\
 &   flat & - & - &  307.5 $\pm$   19.3 &   15.9 &   18.6 $\pm$    6.5 &   1.88 $\pm$   0.17 &     45 &  42.87 & 5.63e-01\\
 &      - & - & - &    0.0 & - &  283.6 $\pm$   14.6 &   0.58 $\pm$   0.04 &     46 & 157.03 & 4.77e-14\\
2A 335+096 &   extr &     37 &   0.12 &    5.3 $\pm$    0.2 &   34.8 &  137.7 $\pm$    1.9 &   1.43 $\pm$   0.02 &     34 & 173.51 & 1.26e-20\\
 &      - & - & - &    0.0 & - &  117.7 $\pm$    1.5 &   1.06 $\pm$   0.01 &     35 & 1188.38 & 6.24e-227\\
 &   flat & - & - &    7.1 $\pm$    0.1 &   49.3 &  138.6 $\pm$    1.9 &   1.52 $\pm$   0.02 &     34 & 209.16 & 4.39e-27\\
 &      - & - & - &    0.0 & - &  107.4 $\pm$    1.4 &   0.97 $\pm$   0.01 &     35 & 2097.26 & 0.00e+00\\
\enddata
\tablecomments{Col. (1) Cluster name; col. (2) CDA observation identification number; col. (3) method of $T_X$ interpolation (discussed in \S\ref{sec:kpr}); col. (4) maximum radius for fit; col. (5) number of radial bins included in fit; col. (6) best-fit core entropy; col. (7) number of sigma \kna\ is away from zero; col. (9) best-fit entropy at 100 kpc; col. (10) best-fit power-law index; col. (11) degrees of freedom in fit; col. (12) \chisq\ statistic of best-fit model; and col. (13) probability of worse fit given \chisq\ and degrees of freedom.}
\end{deluxetable}

%% file: figs.tex
\clearpage
\begin{figure}[htp]
  \begin{center}
    \begin{minipage}[htp]{0.9\linewidth}
      \includegraphics*[width=\textwidth, trim=15mm 10mm 10mm 10mm, clip]{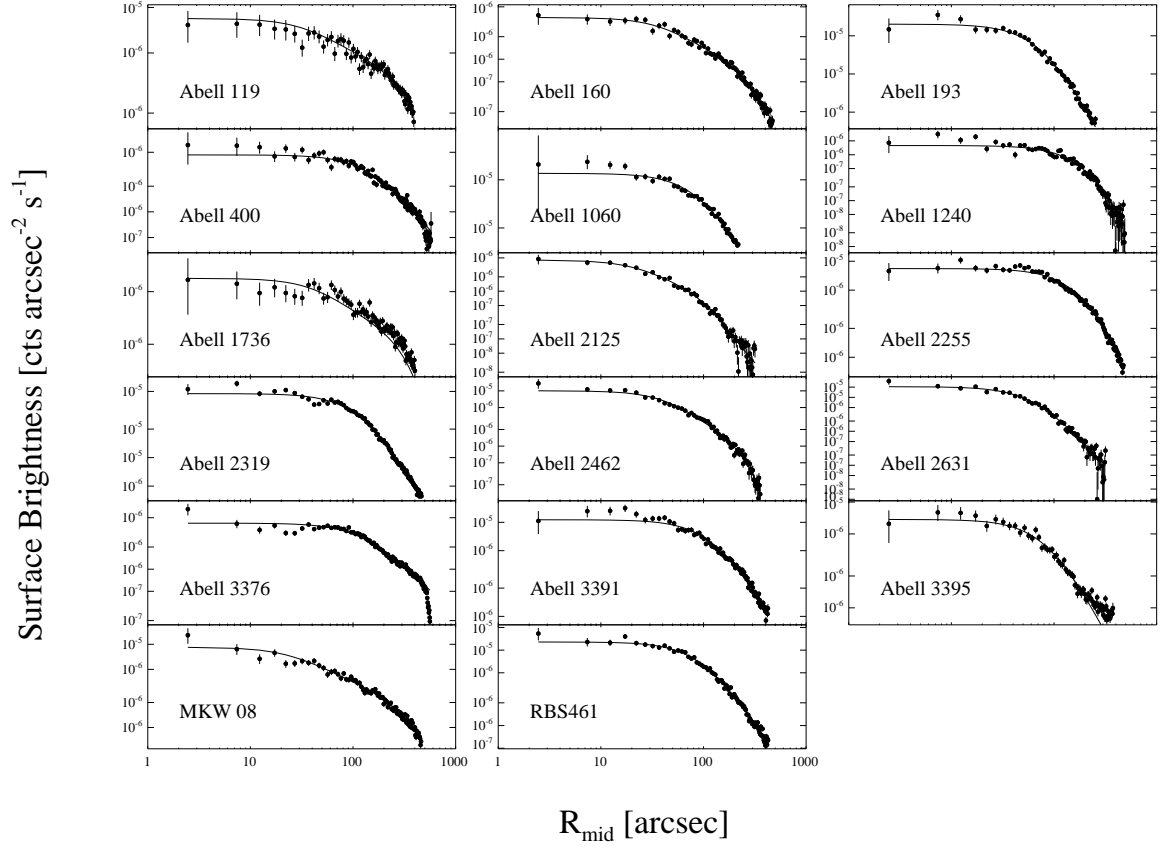}
      \caption{Surface brightness profiles for clusters requiring a
        $\beta$-model fit for deprojection (discussed in
        \S\ref{sec:beta}). The best-fit $\beta$-model for each cluster
        is overplotted as a dashed line. The discrepancy between the
        data and best-fit model for some clusters results from the
        presence of a compact X-ray source at the center of the
        cluster. These cases are discussed in Appendix
        \ref{app:beta}.}
      \label{fig:betamods}
    \end{minipage}
  \end{center}
\end{figure}
\clearpage
\begin{figure}[htp]
  \begin{center}
    \begin{minipage}[htp]{0.9\linewidth}
      \includegraphics*[width=\textwidth, trim=5mm 0mm 5mm 5mm, clip]{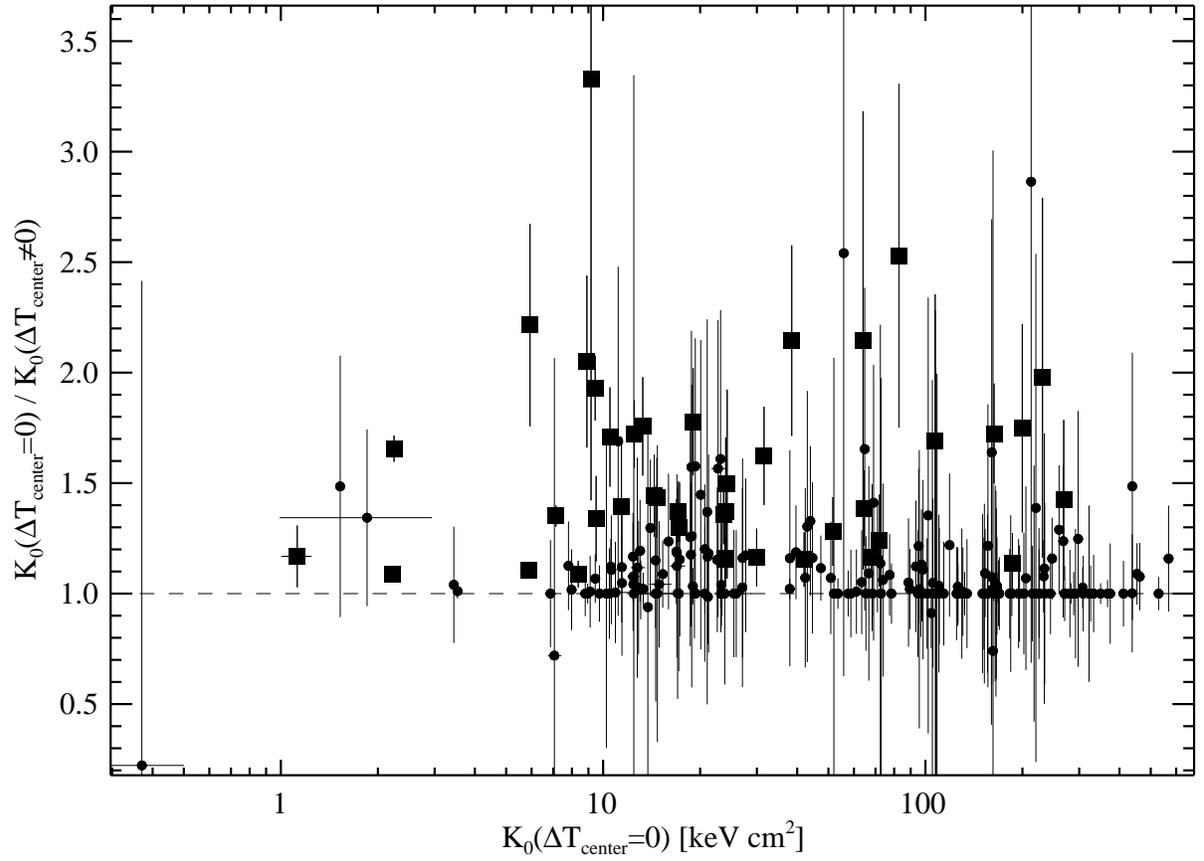}
      \caption{Ratio of best-fit \kna\ for the two treatments of
        central temperature interpolation (see \S\ref{sec:temppr}):
        (1) temperature is free to decline across the central density
        bins ($\Delta T_{center} \ne 0$), and (2) the temperature
        across the central density bins is isothermal ($\Delta
        T_{center} = 0$). Filled black squares are clusters for which
        the \kna\ ratio is inconsistent with unity.}
      \label{fig:kcomp}
    \end{minipage}
  \end{center}
\end{figure}
\clearpage
\begin{figure}[htp]
  \begin{center}
    \begin{minipage}[htp]{0.9\linewidth}
      \includegraphics*[width=\textwidth, trim=5mm 0mm 5mm 5mm, clip]{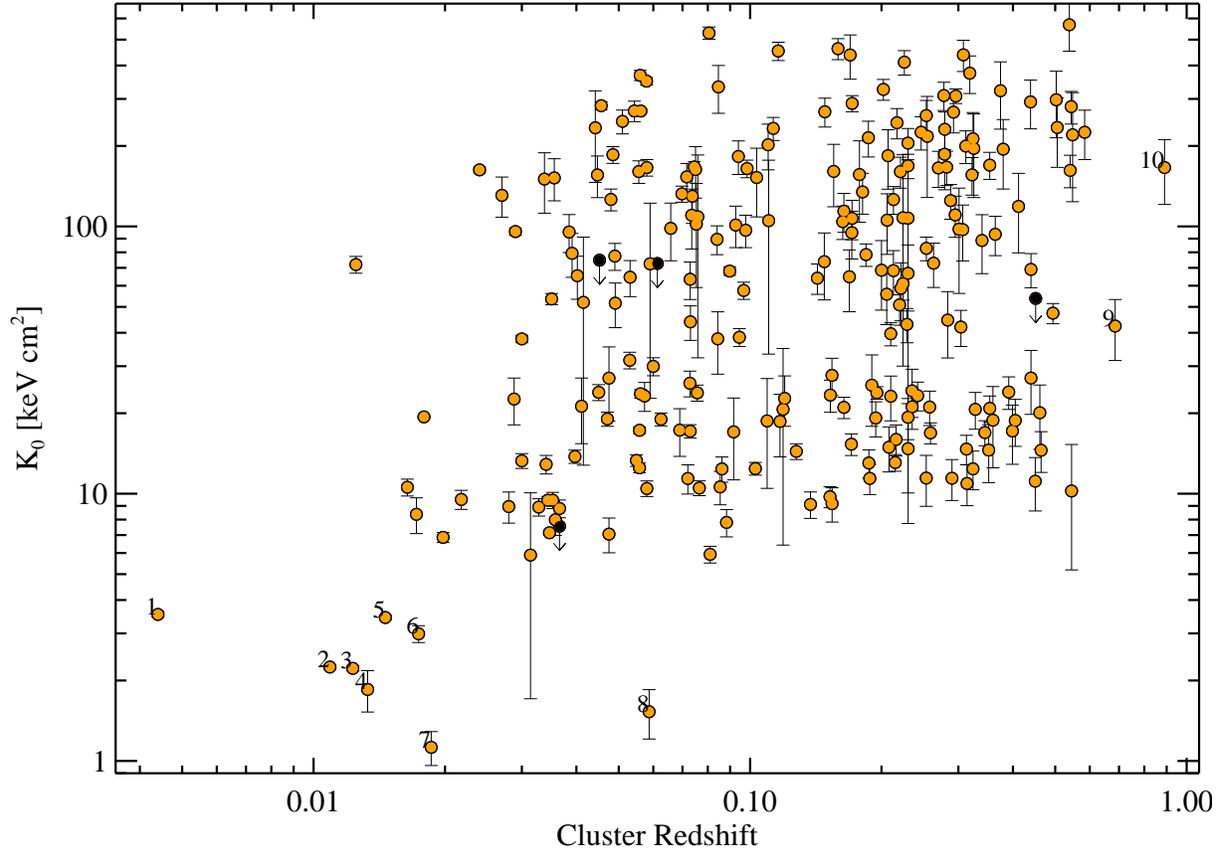}
      \caption{Best-fit \kna\ vs. redshift. Some clusters have
        \kna\ error bars smaller than the point. The clusters with
        upper-limits ({\it{black points with downward arrows}}) are:
        A2151, AS0405, MS 0116.3-0115, and RX J1347.5-1145. The
        numerically labeled clusters are: (1) M87, (2) Centaurus
        Cluster, (3) RBS 533, (4) HCG 42, (5) HCG 62, (6) SS2B153, (7)
        A1991, (8) MACS0744.8+3927, and (9) CL J1226.9+3332. For
        CLJ1226, \cite{2007ApJ...659.1125M} found best-fit $\kna = 132
        \pm 24 \ent$ which is not significantly different from our
        value of $\kna = 166 \pm 45 \ent$. The lack of $\kna < 10
        \ent$ clusters at $z > 0.1$ is most likely the result of
        insufficient angular resolution (see \S\ref{sec:angres}).}
      \label{fig:k0res}
    \end{minipage}
  \end{center}
\end{figure}
\clearpage
\begin{center}
  \begin{figure}[htp]
    \begin{minipage}[htp]{0.5\linewidth}
      \includegraphics*[width=\textwidth, trim=28mm 7mm 30mm 17mm, clip]{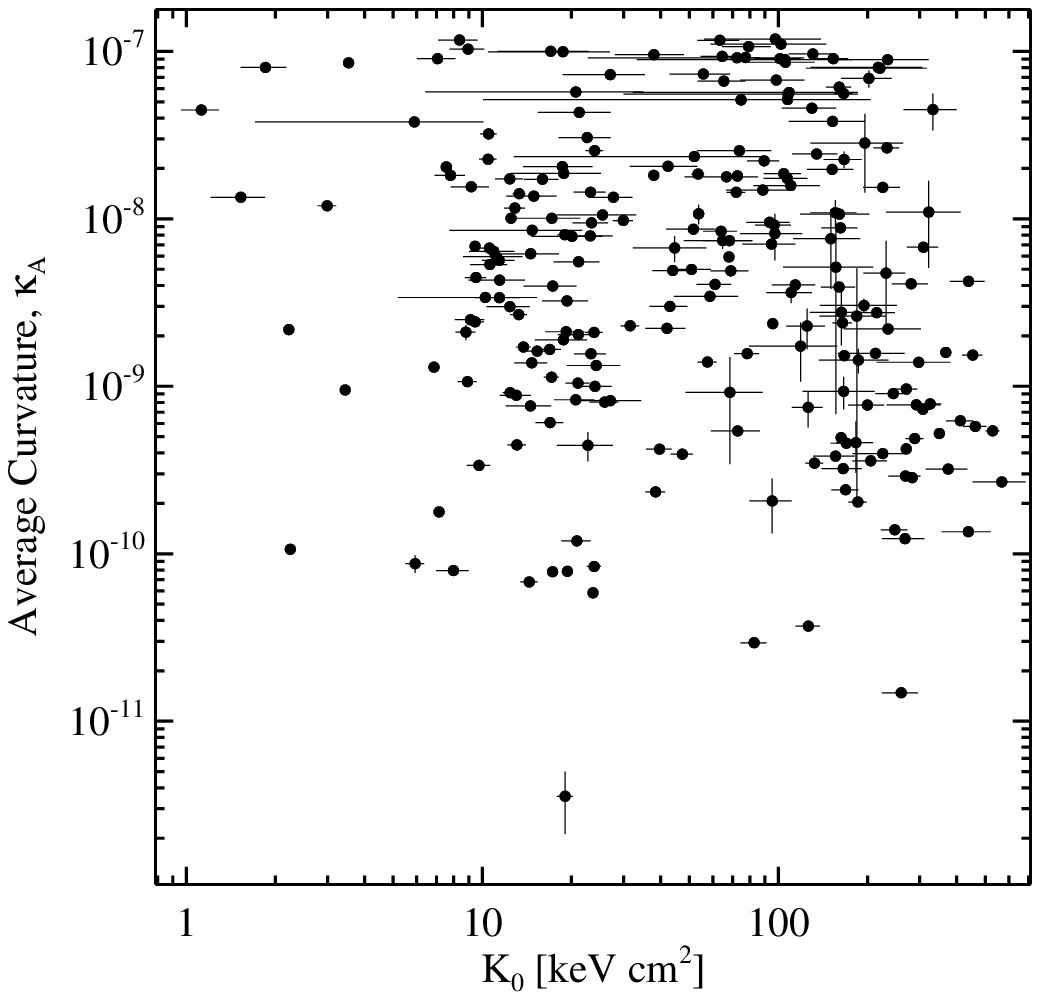}
    \end{minipage}
    \begin{minipage}[htp]{0.5\linewidth}
      \includegraphics*[width=\textwidth, trim=28mm 7mm 30mm 17mm, clip]{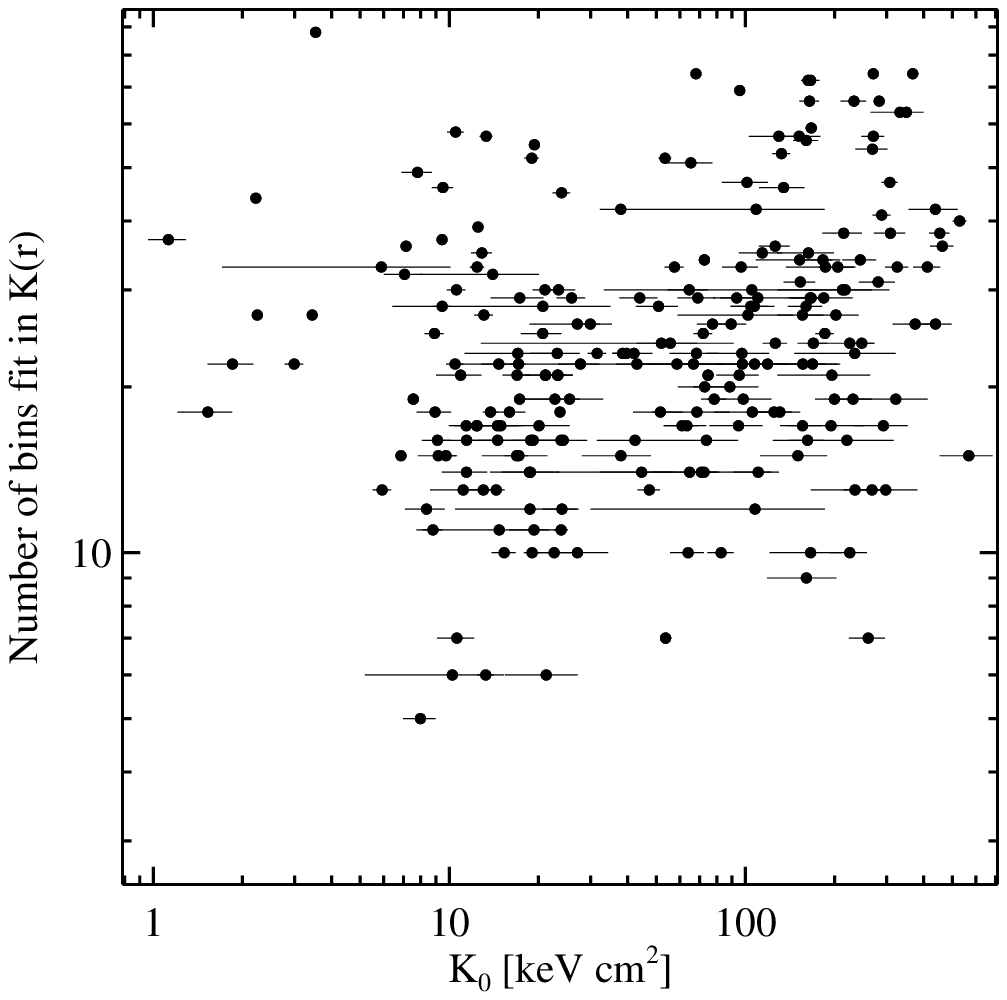}
    \end{minipage}
    \begin{minipage}[htp]{0.5\linewidth}
      \includegraphics*[width=\textwidth, trim=28mm 7mm 30mm 17mm, clip]{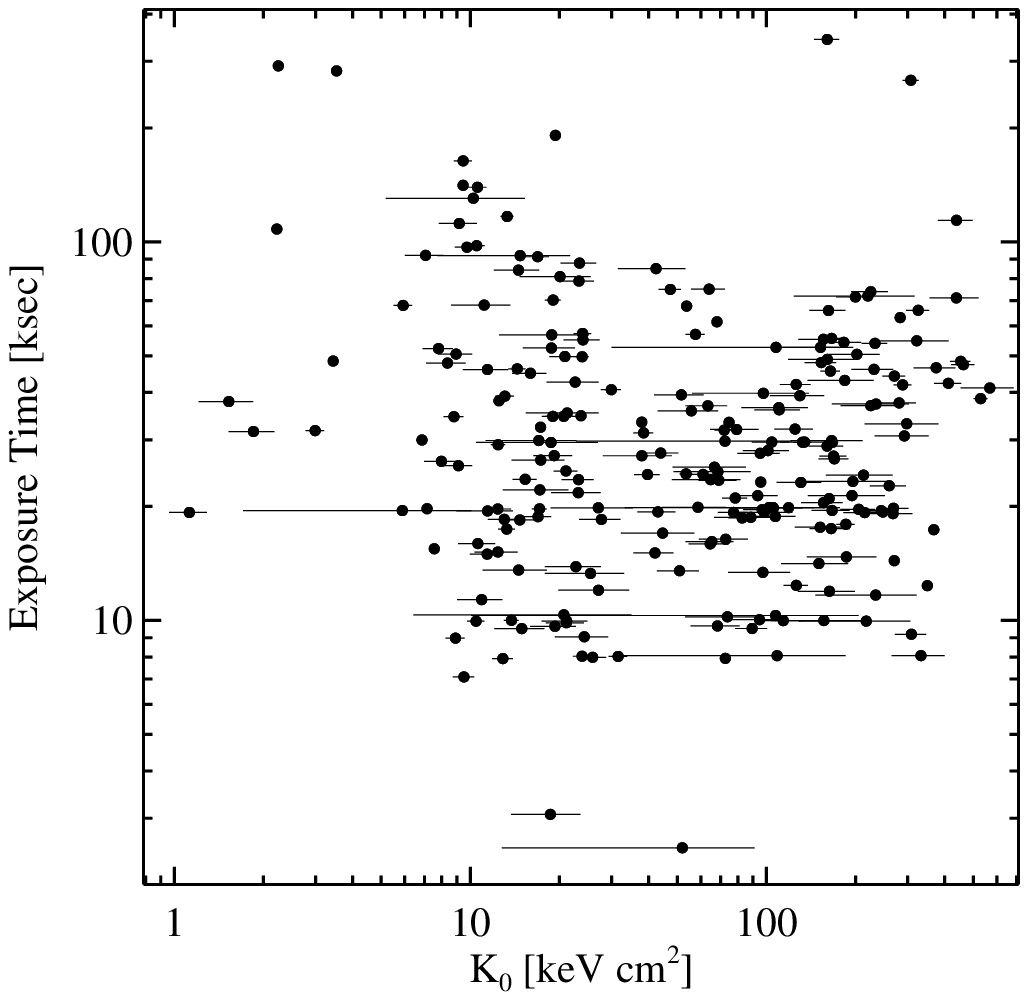}
    \end{minipage}
    \begin{minipage}[htp]{0.5\linewidth}
      \includegraphics*[width=\textwidth, trim=28mm 7mm 30mm 17mm, clip]{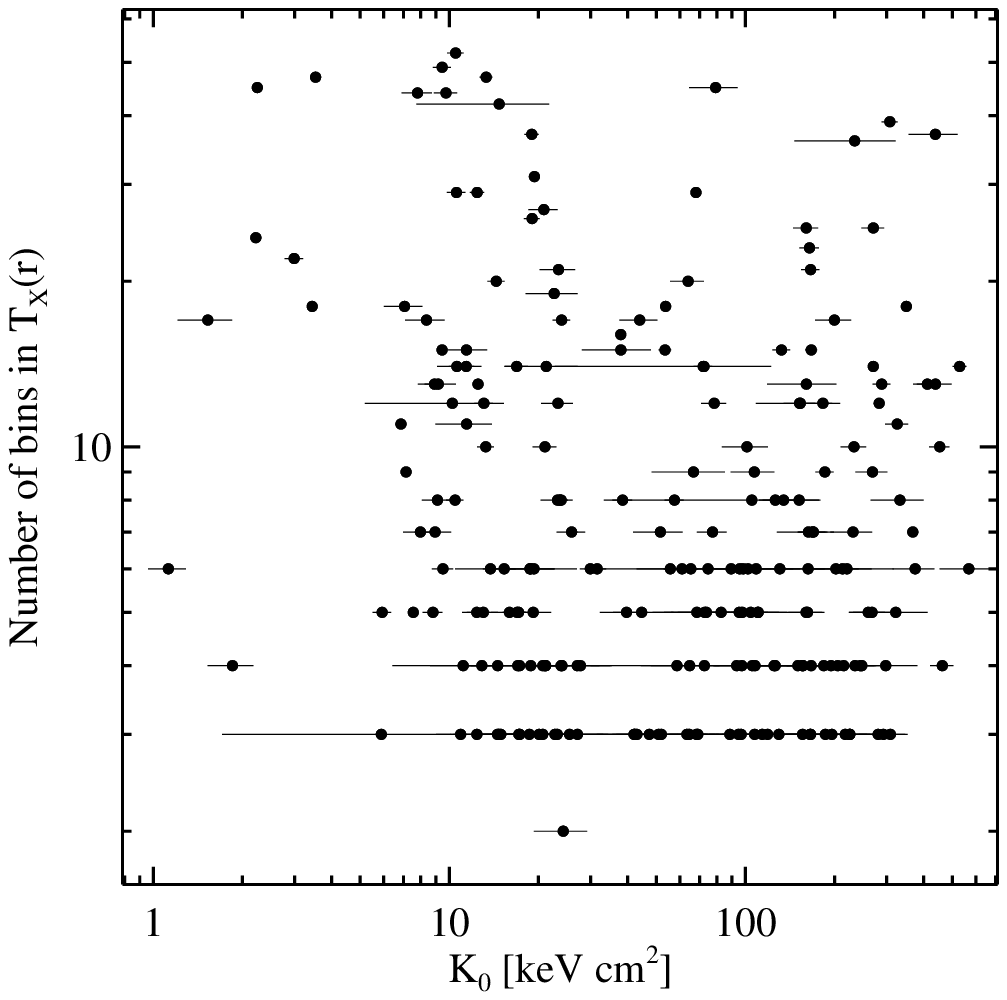}
    \end{minipage}
    \caption{Plots of possible systematics versus best-fit \kna.
      {\it{Top left:}} Best-fit \kna\ plotted versus average curvature
      of the corresponding entropy profile (see eq. \ref{eqn:avgcurv})
      There is no trend between these two quantities suggesting that
      \kna\ is not heavily influenced by the total shape of the
      entropy profile. {\it{Top right:}} Best-fit \kna\ plotted versus
      number of bins in the entropy profile which were used during
      fitting. Again, no trend is found. {\it{Bottom left:}} Best-fit
      \kna\ plotted versus the total used exposure time for each
      cluster. No trend is found. {\it{Bottom right:}} Best-fit
      \kna\ plotted versus the number of bins in the temperature
      profile for each cluster. As expected, fewer $\Tx(r)$ does not
      correlate with \kna.}
    \label{fig:sys}
  \end{figure}
\end{center}
\clearpage
\begin{center}
  \begin{figure}[htp]
    \begin{minipage}[htp]{0.5\linewidth}
      \includegraphics*[width=\textwidth, trim=28mm 7mm 30mm 17mm, clip]{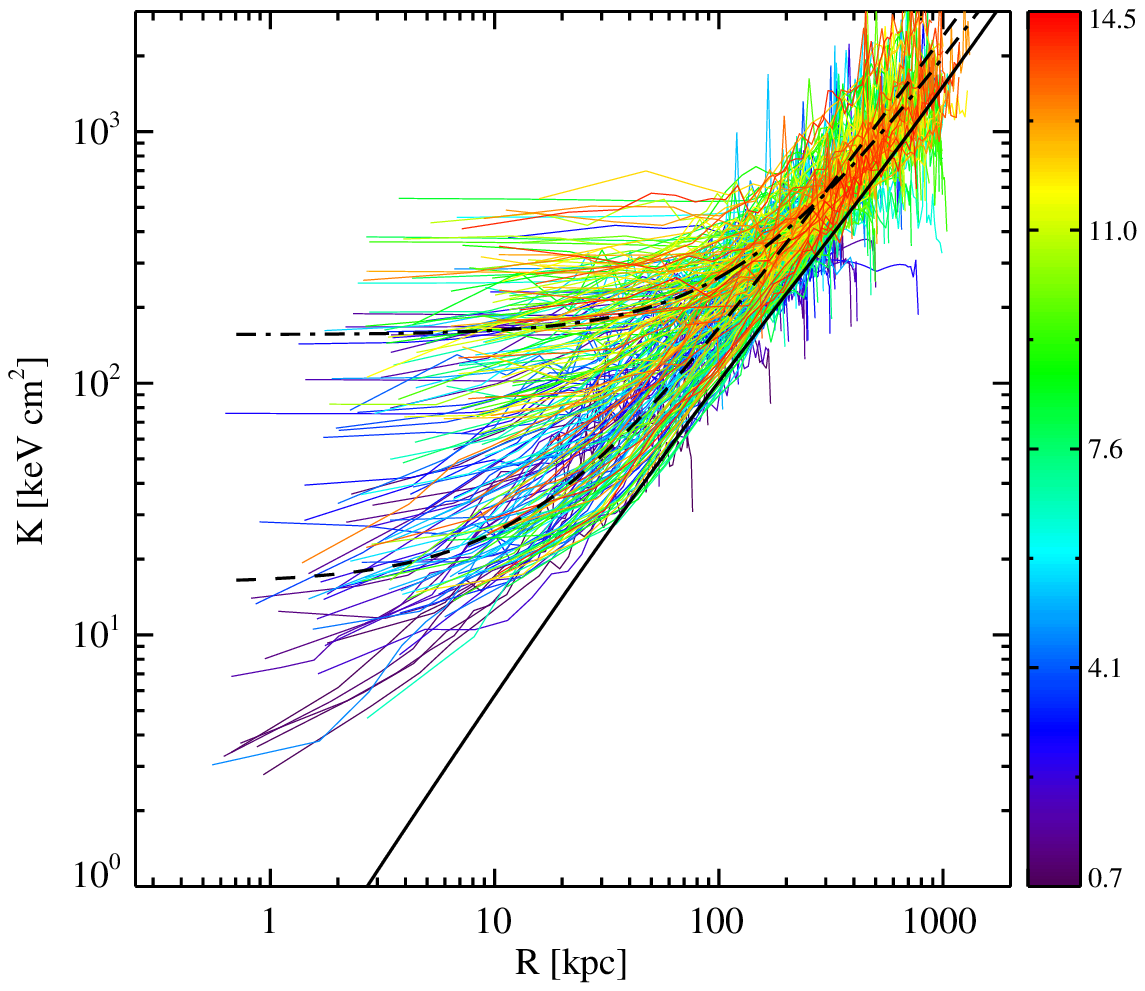}
    \end{minipage}
    \begin{minipage}[htp]{0.5\linewidth}
      \includegraphics*[width=\textwidth, trim=28mm 7mm 30mm 17mm, clip]{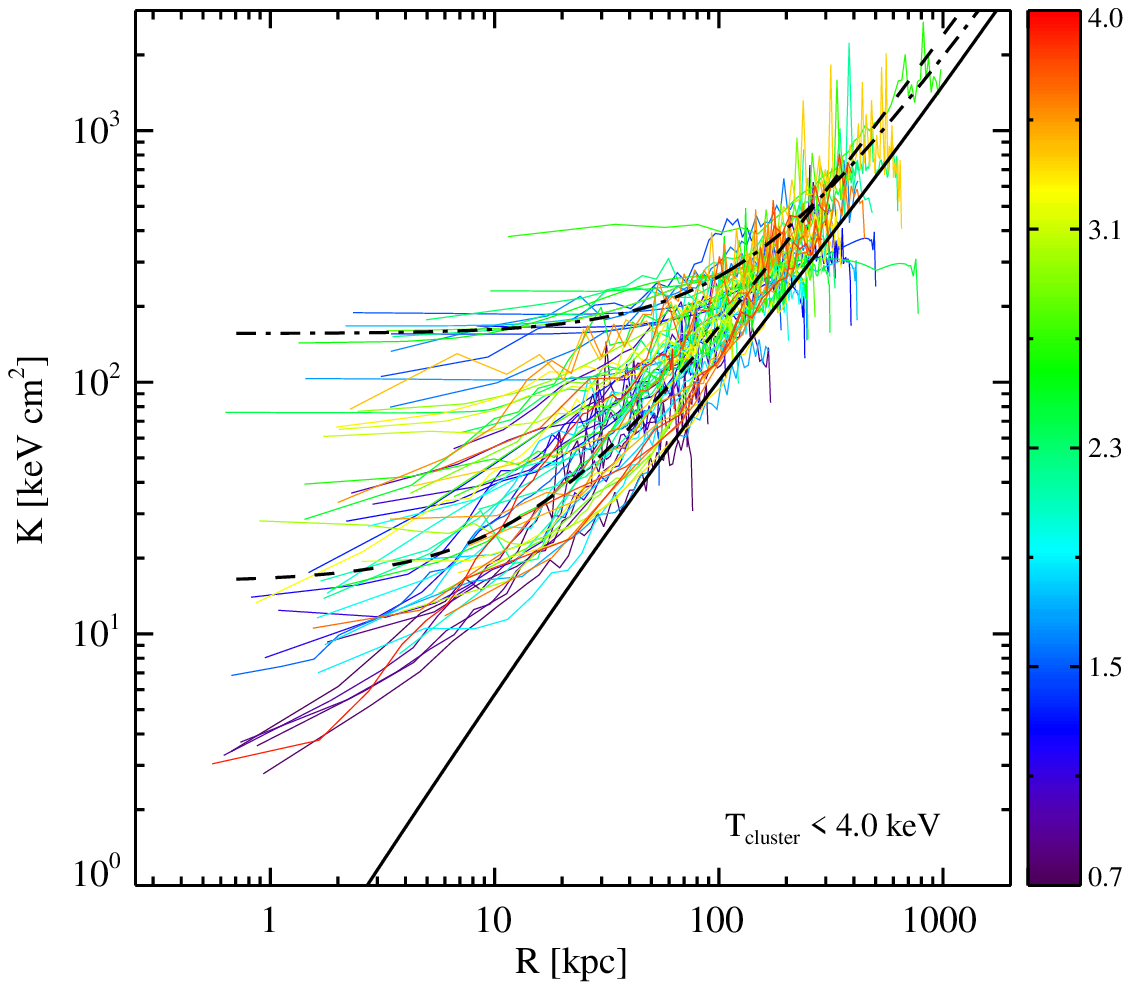}
    \end{minipage}
    \begin{minipage}[htp]{0.5\linewidth}
      \includegraphics*[width=\textwidth, trim=28mm 7mm 30mm 17mm, clip]{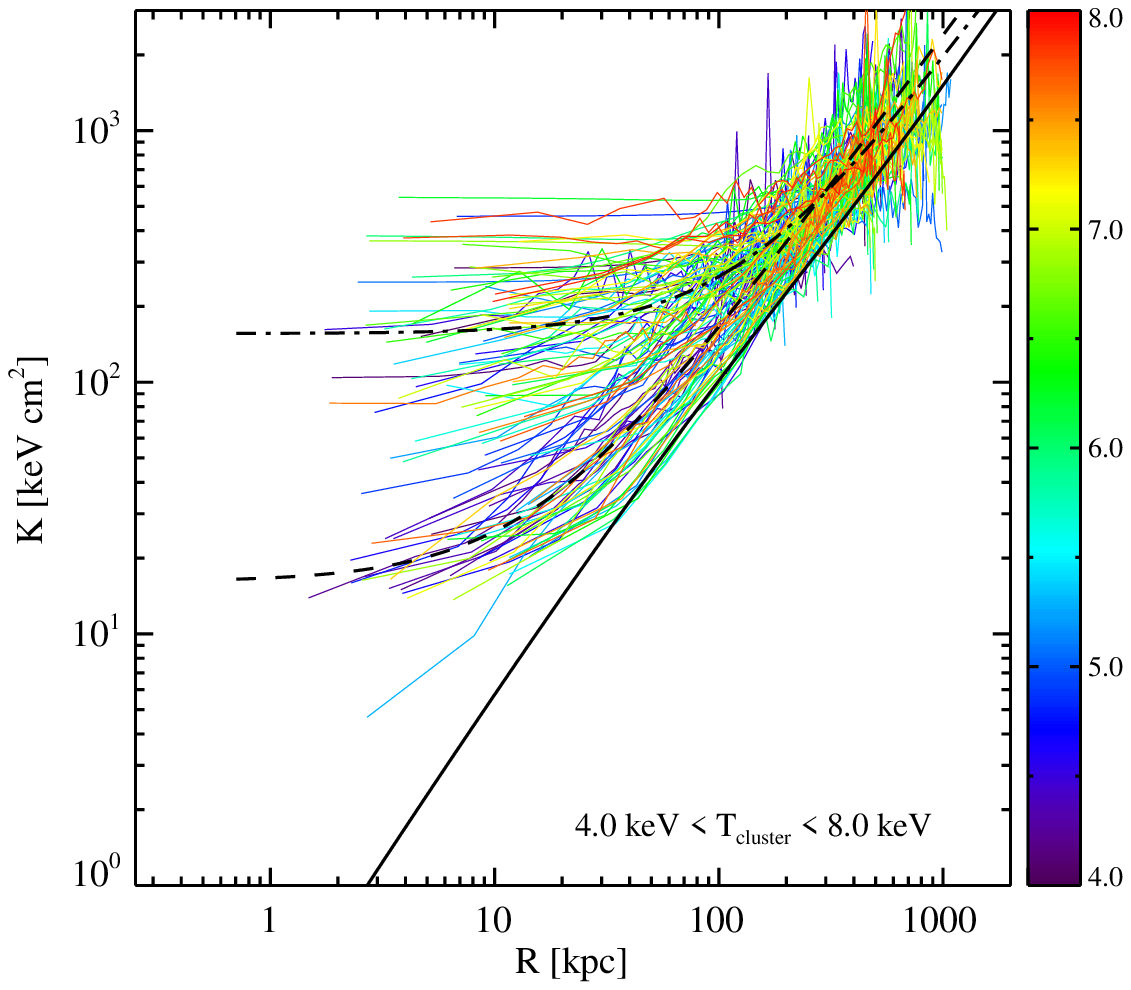}
    \end{minipage}
    \begin{minipage}[htp]{0.5\linewidth}
      \includegraphics*[width=\textwidth, trim=28mm 7mm 30mm 17mm, clip]{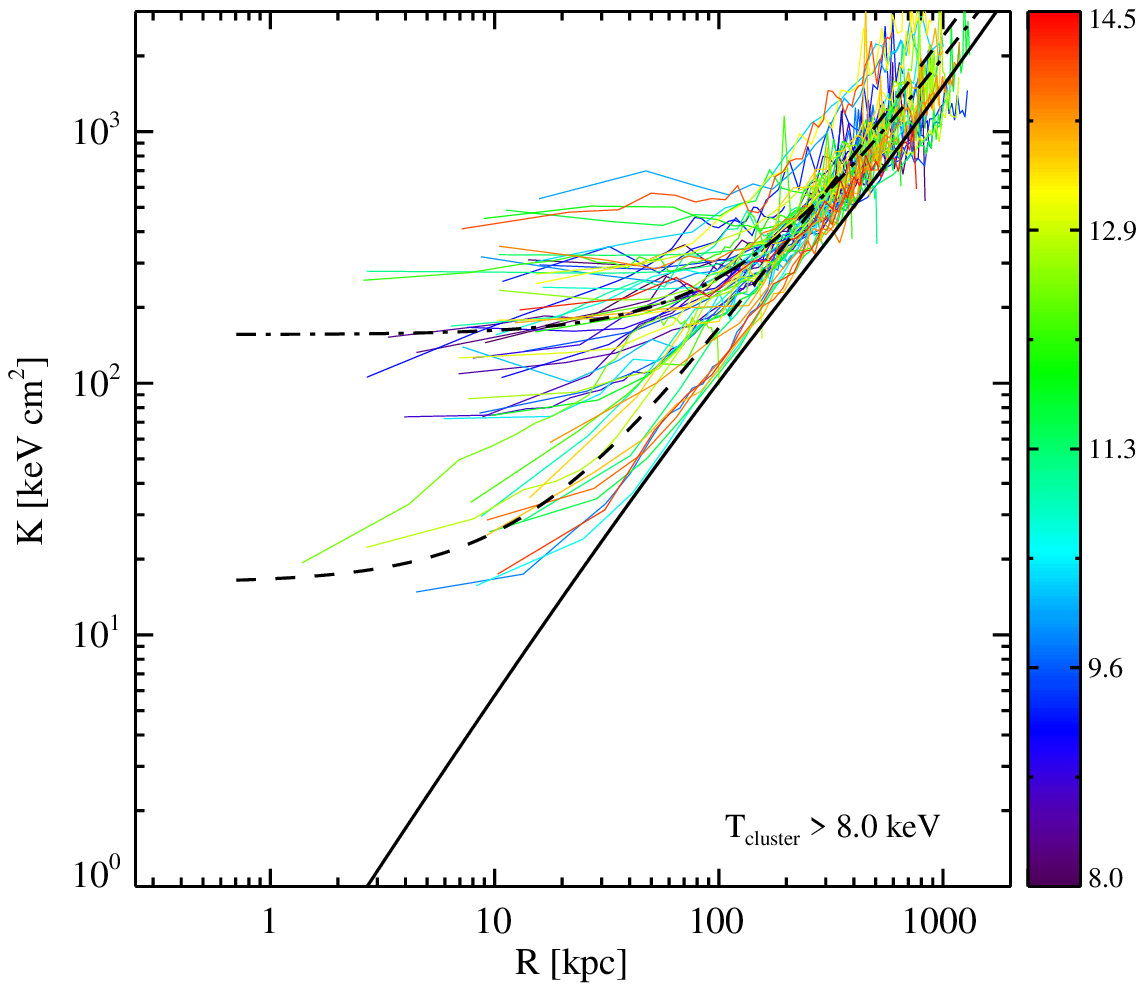}
    \end{minipage}
    \caption{Composite plots of entropy profiles for varying cluster
      temperature ranges. Profiles are color-coded based on average
      cluster temperature. Units of the color bars are keV. The solid
      line is the pure-cooling model of \cite{voitbryan}, the dashed
      line is the mean profile for clusters with $\kna \le 50 \ent$,
      and the dashed-dotted line is the mean profile for clusters with
      $\kna > 50 \ent$. {\it{Top left:}} This panel contains all the
      entropy profiles in our study. {\it{Top right:}} Clusters with
      $kT_X < 4$ keV. {\it{Bottom left:}} Clusters with $4\keV < kT_X
      < 8\keV$. {\it{Bottom right:}} Clusters with $kT_X > 8$
      keV. Note that while the dispersion of core entropy for each
      temperature range is large, as the $kT_X$ range increases so to
      does the mean core entropy.}
    \label{fig:splots}
  \end{figure}
\end{center}
\clearpage
\begin{figure}[htp]
  \begin{center}
    \begin{minipage}[htp]{0.9\linewidth}
      \includegraphics*[width=\textwidth, trim=20mm 10mm 10mm 10mm, clip]{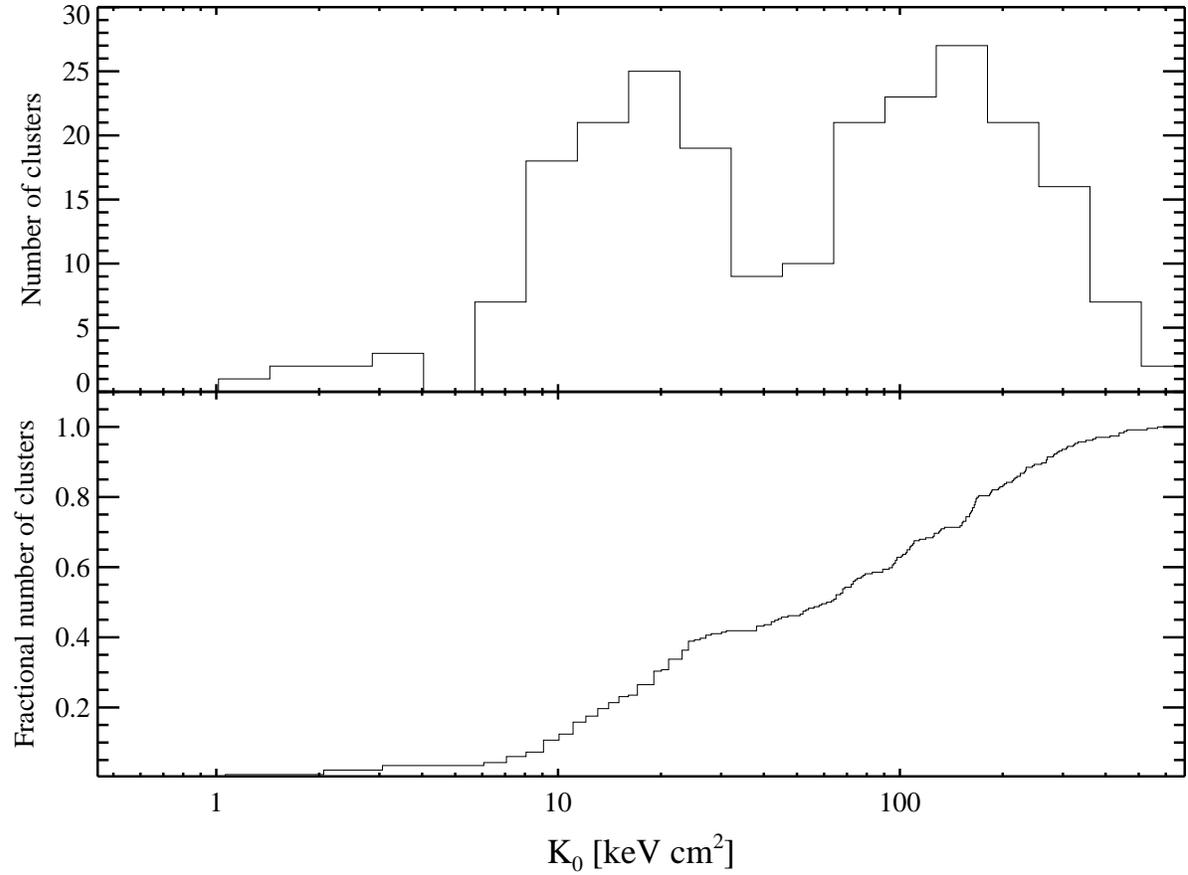}
      \caption{{\it{Top panel:}} Histogram of best-fit \kna\ for all
        the clusters in \accept. Bin widths are 0.15 in log space.
        {\it{Bottom panel:}} Cumulative distribution of \kna\ values
        for the full sample. The distinct bimodality in \kna\ is
        present in both distributions, which would not be seen if it
        were an artifact of the histogram binning. A KMM test finds
        the \kna\ distribution cannot arise from a simple unimodal
        Gaussian.}
      \label{fig:k0hist}
    \end{minipage}
  \end{center}
\end{figure}
\clearpage
\begin{figure}[htp]
  \begin{center}
    \begin{minipage}[htp]{0.9\linewidth}
      \includegraphics*[width=\textwidth, trim=20mm 10mm 10mm 10mm, clip]{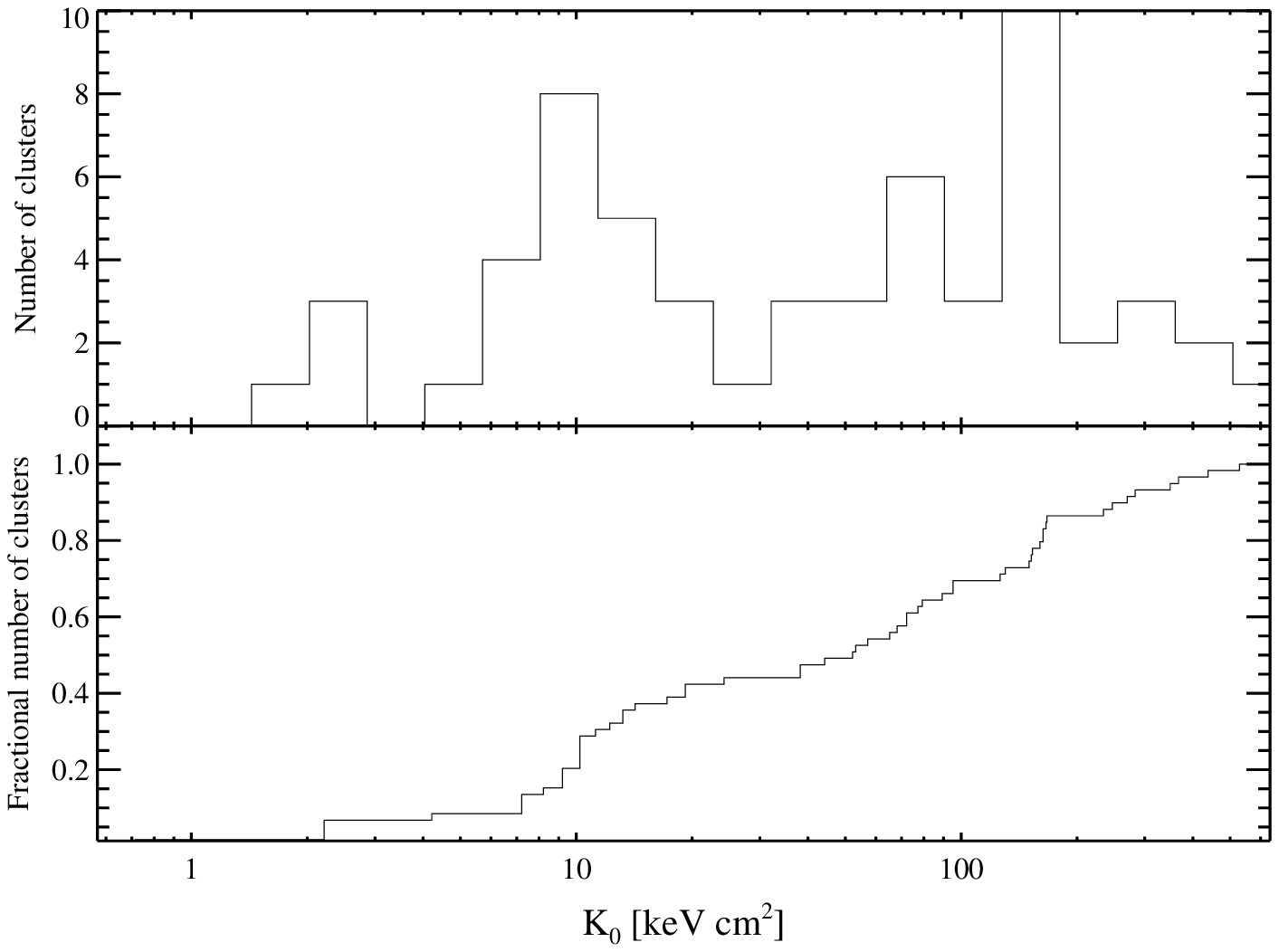}
      \caption{{\it{Top panel:}} Histogram of best-fit \kna\ values
        for the primary \hifl\ sample. Bin widths are 0.15 in log
        space.  {\it{Bottom panel:}} Cumulative distribution of
        best-fit \kna\ values. The distinct bimodality seen in the
        full \accept\ sample (Fig. \ref{fig:k0hist}) is also present
        in the \hifl\ subsample and shares the same gap between the
        low-entropy peak at 10-20 \ent\ and the high-entropy peak at
        100-200 \ent. That bimodality is present in both samples is
        strong evidence it is not a result of an unknown archival
        bias.}
      \label{fig:hiflk0}
    \end{minipage}
  \end{center}
\end{figure}
\clearpage
\begin{figure}[htp]
  \begin{center}
    \begin{minipage}[htp]{0.8\linewidth}
      \includegraphics*[width=\textwidth, trim=20mm 10mm 10mm 10mm, clip]{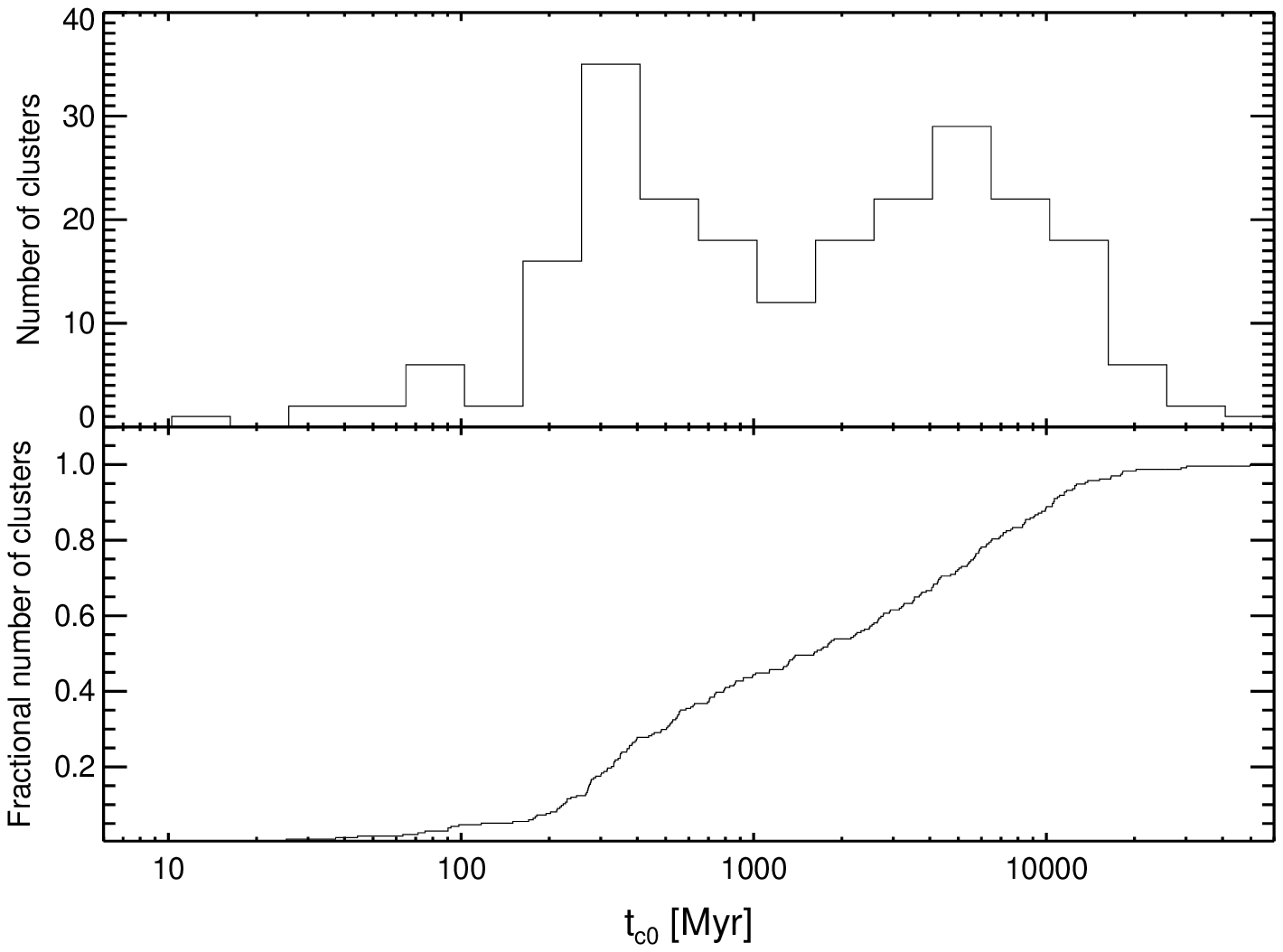}
    \end{minipage}
    \begin{minipage}[htp]{0.8\linewidth}
      \includegraphics*[width=\textwidth, trim=20mm 10mm 10mm 10mm, clip]{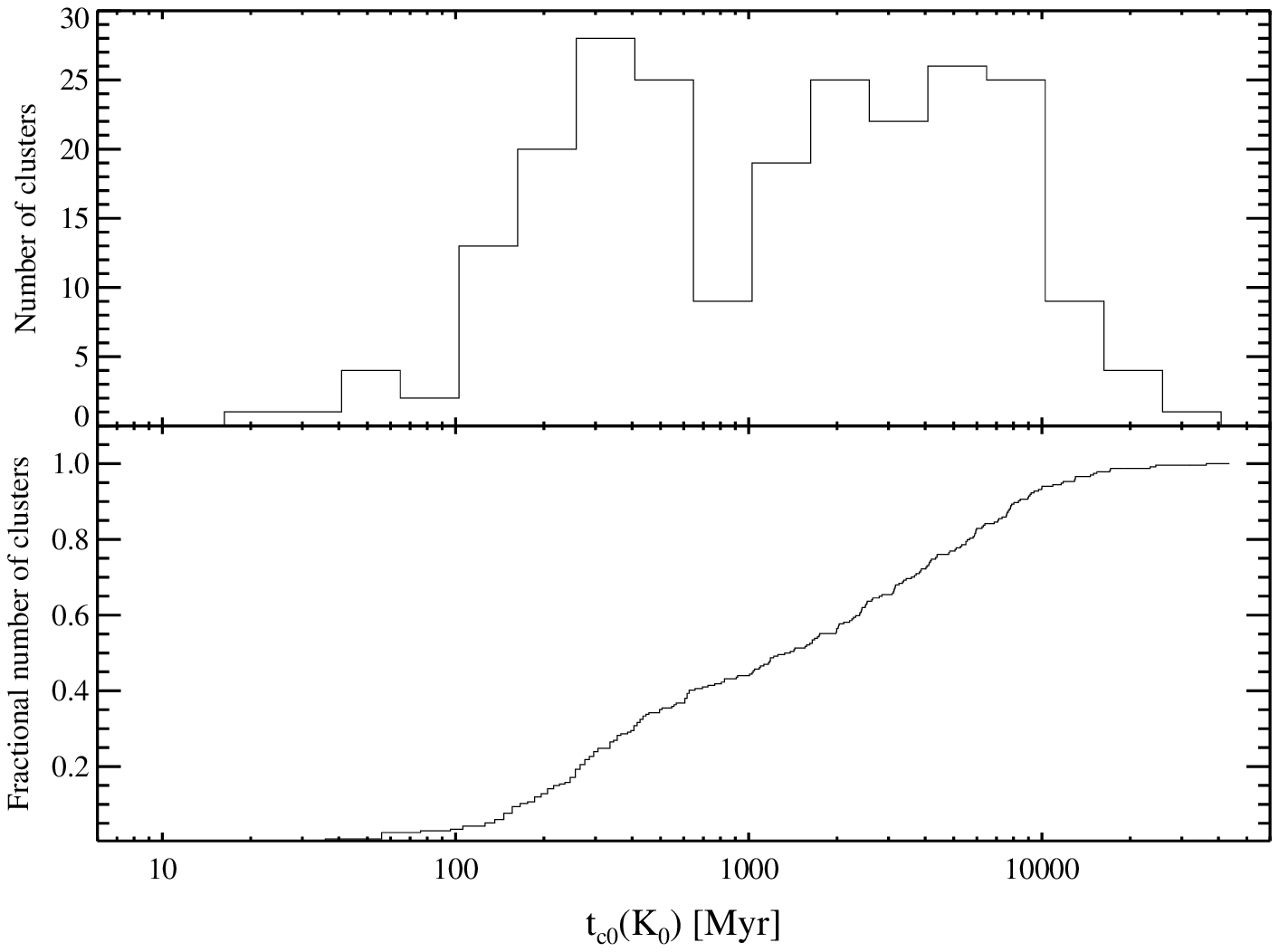}
    \end{minipage}
    \caption{{\it{Top panel:}} Log-binned histogram and cumulative
      distribution of best-fit core cooling times, $t_{c0}$
      (eqn. \ref{eqn:tc0}), for all the clusters in \accept. Histogram
      bin widths are 0.2 in log space. {\it{Bottom panel:}} Log-binned
      histogram and cumulative distribution of core cooling times
      calculated from best-fit \kna\ values, $t_{c0}(\kna)$
      (eqn. \ref{eqn:tck0}), for all the clusters in
      \accept. Histogram bin widths are 0.2 in log space. The
      bimodality we observe in the \kna\ distribution is also present
      in best-fit $t_{c0}$. However, the gaps between the two
      populations of $t_{c0}$ and $t_{c0}(\kna)$ differ by $\sim 0.3$
      Gyrs which may be an artifact of the binning.}
    \label{fig:t0}
  \end{center}
\end{figure}